\documentclass[twoside,12pt]{article}
\usepackage{macros2e}
\usepackage{graphicx}
\usepackage{amssymb}
\usepackage{version}
\usepackage{amsmath}
\usepackage{bildmachen}
\usepackage{times}

\newcommand{\rme}{{\mathrm{e}}}
\newcommand{\rmd}{{\mathrm{d}}}
\newcommand {\ds}{\displaystyle}

\newcommand{\nn}{\nonumber}

\newcommand{\Tr}{{\mathrm{Tr}}} 
\begin{document}
\begin{titlepage}
\noindent
\renewcommand{\thefootnote}{\fnsymbol{footnote}}
\vfill 
\centerline{\sffamily\bfseries\Large Scaling behavior of tethered crumpled}
\medskip
\centerline{\sffamily\bfseries\Large  manifolds with
inner dimension close to $D=2$:}
\medskip  
\centerline{\sffamily\bfseries\Large Resumming the perturbation theory.} 
\vfill
\centerline{\bf\large Henryk A.\ Pinnow$^{1}$ and Kay J\"org Wiese$^{1,2}$%
\footnote{Email: hpinnow@sinits.com, wiese@lpt.ens.fr}}
\smallskip
\centerline{\small $^{1}$ Fachbereich Physik, Universit\"at Essen,  45117 Essen,
Germany}\centerline{\small $^{2}$ Laboratoire de Physique Th\'eorique,
Ecole Normale Sup\'erieure, 24 rue Lhomond, 75005 Paris, France}
\smallskip\smallskip

\vfill
\vspace{-5mm}
\begin{abstract}\noindent The field theory of self-avoiding tethered
membranes still poses major challenges.  In this article, we report
progress on the toy-model of a manifold repelled by a single point.
Our approach allows to sum the perturbation expansion in the strength
$g_{0}$ of the interaction {\em exactly} in the limit of internal
dimension $D\to 2$, yielding an analytic solution for the
strong-coupling limit.  This analytic solution is the starting point
for an expansion in $2-D$, which aims at connecting to the well
studied case of polymers ($D=1$). We give results to fourth order in
$2-D$, where the dependence on $g_{0}$ is again summed exactly.  As an
application, we discuss plaquette density functions, and propose a
Monte-Carlo experiment to test our results.  These methods should also
allow to shed light on the more complex problem of self-avoiding
manifolds.


\end{abstract}
\vspace{-5mm}
\vfill

\centerline{\em Submitted to Nuclear Physics B} 
\vfill
\vspace*{3cm}

%
%
\end{titlepage}

\renewcommand{\thefootnote}{\fnsymbol{footnote}}
{
\setcounter{tocdepth}{4}
\tableofcontents} \newpage
\setcounter{page}{3}

\renewcommand{\thefootnote}{\arabic{footnote}}

\savebild{\GA}{\bildGA}{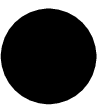}
\savebild{\GB}{\bildGB}{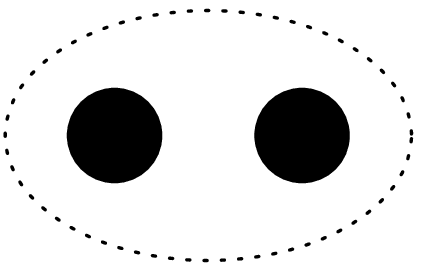}
\savebild{\GC}{\bildGC}{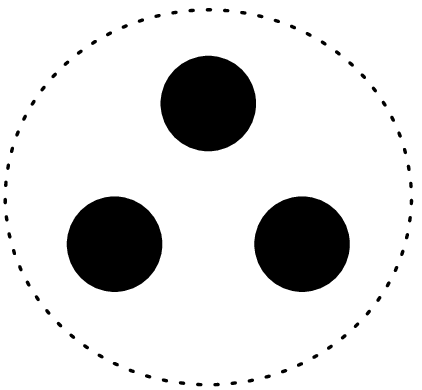}
\savebild{\GD}{\bildGD}{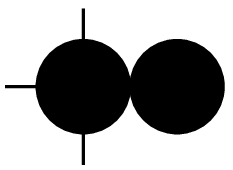}
\savebild{\BT}{\bildBT}{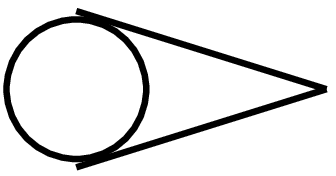}

\newcommand{\C}{\mathbb{C}}
\newcommand{\ovC}{\overline{\C}}
\newcommand{\Csq}{\overline{\C}^{2}}
\newcommand{\Ccsq}{\overline{\C_{c}^{2}}}
\newcommand{\Cccub}{\overline{\C_{c}^{3}}}
\newcommand{\Cctri}{\overline{\C_{c}^{\triangle}}}
\newcommand{\Ccdia}{\overline{\C_{c}^{\diamond}}}
\newcommand{\Ccbisq}{\overline{\C_{c}^{4}}}
\newcommand{\Ccblobt}{\overline{\C_{c}^{\BT}}}
\newcommand{\mathematica}{{\em Mathematica$^{\bigcirc\hspace{-2.3mm}\mbox{\tiny R}}$\/}}

\setcounter{footnote}{0}

\section{Introduction}\label{introduction} One major problem in
statistical physics is the effect of interactions on the
thermodynamical properties of extended fluctuating geometric
objects. In general, multi-particle attractive or repulsive
interactions are involved. One may divide these into two classes:
Either (i) one may study the interaction of a single fluctuating
object with itself as for instance the well known excluded volume
interaction between any two monomers in a long polymer chain in a good
solvent.  There, the interaction leads to universal long-distance
properties of chains as for example the anomalous scaling of the mean
squared end-to-end distance. Or (ii), the interaction may act between
different manifolds or between a single manifold and a fixed
non-fluctuating object. It is then interesting to study how thermal
fluctuations affect the depinning of the manifold from an attractive
substrate as well as the steric
 repulsions from a wall. Finally, both cases can appear together.\\
 Whatever the situation is, it is usually well understood as long as
the fluctuating objects are one-dimensional
\cite{Schaefer,DesCloizeauxJannink,DeGennes,Eisenriegler}. Referring
to the example mentioned above, the long-distance properties of
self-avoiding polymers can be analyzed with renormalization group
techniques \cite{Fixman1955,SchaferWitten1977,DesCloizeaux1981},
either in the continuous Edwards Hamiltonian \cite{Edwards1965},
\begin{equation}
  \label{I.0}
  \mathcal{H}[\vec r]=\frac{1}{2}\int\limits_{x\in\mathcal{M}}\left(\nabla
  \vec
  r(x)\right)^{2}+\frac{b_{0}}{2}\int\limits_{x\in\mathcal{M}}\int\limits_{y\in\mathcal{M}}\delta^{d}(\vec
  r(x)-\vec r(y))\ ,
\end{equation}
or by mapping this model on a local $O(N)$ symmetric
$\varphi^{4}$-theory in the limit of $N=0$ components
\cite{DeGennes1972,DeGennes,Schaefer}. The critical exponents
describing the long-distance properties are related to the critical
exponents of the corresponding $N$-vector model at the critical
point. What makes (\ref{I.0}) a non-standard theory is that the
interaction is non-local, and not a polynomial of the field.\\
Obtaining the corresponding results for membranes poses considerable
challenges. The generalization of polymers to $2D$-surfaces are
crystalline fixed-connectivity membranes as they appear for instance
in the spectrin network of cell membranes. Considering ``phantom''
membranes which can freely fold into itself, the existence of a
bending rigidity induced phase transition separating a high rigidity,
low temperature flat phase from a low rigidity, high temperature
crumpled phase is well established
\cite{KantorNelson1987a,KantorNelson1987b,KantorKardarNelson1986a,KantorKardarNelson1986b,PaczuskiKardarNelson1988,PaczuskiKardar1989}. This
is in contrast to polymers, which are always crumpled on large
scales. The scaling properties of the crumpled phase of phantom
membranes are described by the $2D$ generalization of the free field
part in (\ref{I.0}). Taking self-avoidance into account, which is
modeled in (\ref{I.0}) through the short-range two-body interaction,
we expect more swollen manifolds than those predicted by the free theory,
which will be expressed in a non-trivial radius of gyration exponent
$\nu$:
\begin{equation}
  \label{I.01}
  R_{g}\sim L^{\nu}\ ,\quad 0\leq\nu\leq 1\ ,
\end{equation}
where $L$ denotes the linear internal size of the membrane, and the
radius of gyration $R_{g}$ is obtained from the effective extend of
the membrane in external space.  In the case of polymers $R_{g}$
scales like the end-to-end distance.  Much effort has been spent on
calculating corrections to the radius of gyration exponent within an
expansion in the deviation $\varepsilon$ from the critical space
dimension \cite{DavidWiese1996,WieseDavid1997}. These calculations can
not be performed directly for membrane-dimension $D=2$, since the
naive scaling dimension of the coupling in (\ref{I.0}) equals
\begin{equation}
  \label{I.1a}
  \varepsilon(D,d):=[b_{0}]=2D-\frac{2-D}{2}d\ ,
\end{equation}
where $d$ denotes the dimension of the embedding space, such that
$\epsilon $ is always non-zero as $D\to 2$, for any embedding dimension
$d$. Equivalently, the critical embedding dimension defined through
$\varepsilon(D,d_{c}(D))=0$ becomes infinity in this limit. The reason
is that the {non-self-avoiding} membrane densely fills out the embedding
space, such that it always ``sees'' the interaction. A way to
circumvent this problem is to set up the expansion about any point
$(D<2,d_{c}(D))$ and to extrapolate along an appropriate path in the
$(D,d)$-plane to the physically interesting point $(D,d)=(2,3)$
\cite{WieseHabil,KardarNelson1987,AronovitzLubensky1988,DDG3,DDG4,Hwa1990,WieseDavid1995}. To
second order in $\varepsilon$ one then finds a radius of gyration
exponent of $\nu\approx 0.86$ \cite{DavidWiese1996,WieseDavid1997} ,
which is a strong correction with respect to the only logarithmic
dependence in the non-interacting
theory, and indicates the existence of a crumpled phase, for which
$\nu \ge 2/3$ follows from the fact that a membrane has a finite
volume.\\
However, there is no evidence for a crumpled phase in experiments
\cite{ChianelliPrestridgePecoradoPecoradodDeNeufville1979,HwaKokufutaTanaka1991,WenGarlandHwaKardarKokufutaLiOrkiszTanaka1992,SpectorNaranjoChiruvoluZasadzinski1994}.
Latest Monte-Carlo simulations on plaquette-models
\cite{Baumgaertner1991,BaumgaertnerRenz1992,KrollGompper1993,BowickTravesset2001}
starting from a discretization of the $2D$ generalized Hamiltonian
(\ref{I.0}) with system sizes of up to $\approx 17000$ plaquettes
shows considerable evidence for a vanishing of the above mentioned
crumpling transition in the presence of self-avoidance, such that even
on large scales fixed-connectivity membranes stay always flat with a
radius of gyration exponent of $\nu\approx 1$.
The final purpose of this work is to develop techniques which allow
to go beyond the two-loop result. So far, we developed such
techniques for a simplified model, which reduces the non-local
self-avoiding interaction in (\ref{I.0}) to self-avoidance with only a
single point, e.g.\ the origin $o$ in the membrane:
\begin{eqnarray}
  \label{I.1}
  \mathcal{H}[\vec r]=\frac{1}{2}\int\limits_{x\in\mathcal{M}}\left(\nabla
  \vec
  r(x)\right)^{2}+g_{0}\int\limits_{x\in\mathcal{M}}\delta^{d}(\vec
  r(x)-\vec r(o))\ .
\end{eqnarray}
This is a special case of a more general
model of the interaction of a phantom tethered
manifold with a single point in embedding space, which is related to case
(ii). The corresponding physical situation to think of is the binding and
unbinding
of a long chain as e.g.\ a polymer  or a membrane from a
wall or the wetting of an interface. More precisely, we study 
 the interaction of a single freely
fluctuating manifold with another non-fluctuating, fixed object. 
Depending on whether the interaction is attractive or repulsive, one
can distinguish two different scenarios: One may either observe a
delocalization transition 
 from an
attractive substrate as in wetting phenomena  or
steric repulsions by a fluctuating manifold. Both cases have in common
that excluded volume effects become important. These scenarios have already
been discussed in \cite{PinnowWiese2001}.\\
The result of \cite{PinnowWiese2001} is the complete resummation of
the perturbation series for the effective coupling in the case of
$2D$-membranes. The long-distance behavior of the resummed theory
turned out to be non-trivial in the sense that it emerged from the
limiting behavior of a scale invariant theory resulting in an
effective coupling growing logarithmically instead of approaching some
finite fixed-point value as one would expect it. This, together with
the extremely slow convergence of the perturbation series makes the
analysis of the fully resummed theory a {\em must}, because all
finite loop calculations fail to extract the correct large distance
properties. The importance of the result becomes evident as soon as
one compares it with extrapolations obtained from the
$\varepsilon$-expansion at the 2-loop level
\cite{PinnowWiese2001}. Besides the necessity of calculating diagrams
numerically with considerably raising effort as the loop order becomes
higher, the $\varepsilon$-expansion has turned out not to be able to
make reliable predictions for $D\lesssim 2$. This problem persisted,
though we exploited the freedom to set up the expansion about any
point $(D<2,d_{c}(D))$, $d_{c}(D)=\frac{2D}{2-D}$ being the critical
embedding dimension for given internal dimension $D$ and to expand
both in $D$ and $d$ along any appropriate extrapolation path to some
physically interesting point $(D=2,d)$. As soon as $D$ approaches $2$
the result was
always strongly dependent on the selected expansion point.\\
\\
The aim of this paper is two-fold: First, we reconsider the techniques
to perform loop calculations within a ``massive scheme'', that is on a
manifold of finite size and with fixed space dimension $0<D<2$ and
$d$. We show that the perturbation series for the effective coupling
can be completely summed in $D= 2$ and analyze the long-distance
properties in this limit. In addition to \cite{PinnowWiese2001},
instead of analytically continuing loop integrals to $D=2$ from below
we also perform calculations directly in $2D$, which need an
explicit short-distance (UV)-cutoff. It turns out that results in
$D=2$ are independent of the procedure, i.e.\ they are universal.\\    
Second, we construct a systematic expansion of the effective coupling
in powers of $2-D$. Such an expansion is based on our techniques to
resum the perturbation series at each order in $2-D$. A first attempt
to go beyond $D=2$ has already been made in
\cite{PinnowWiese2001}. However, there, the effect of the boundaries
of the finite manifolds was not taken properly into account, a problem
that has now been circumvented by considering closed manifolds. We
specialize to a toroidal internal topology corresponding to periodic
boundary conditions. Of course, the propagator of the perturbation
series needs to be modified, and diagrams become more difficult to
calculate.  Slightly below $D=2$ we expect power-law behavior of the
effective coupling. We present a possible ansatz for the exact
effective coupling as a function of the internal dimension $D\lesssim
2$, which is consistent with the expansion in $2-D$. However, it
remains an open problem to obtain more information about the power-law
behavior, to make this expansion unique.

A short account of this work has already appeared in
\cite{PinnowWiese2002a}. 

\section{Model and physical observables} \label{Model and physical
observables} \subsection{The model} 
\begin{figure}[htbp]
\begin{minipage}[t]{8cm}
  \begin{center}
    \raisebox{0mm}{\includegraphics[scale=0.4]{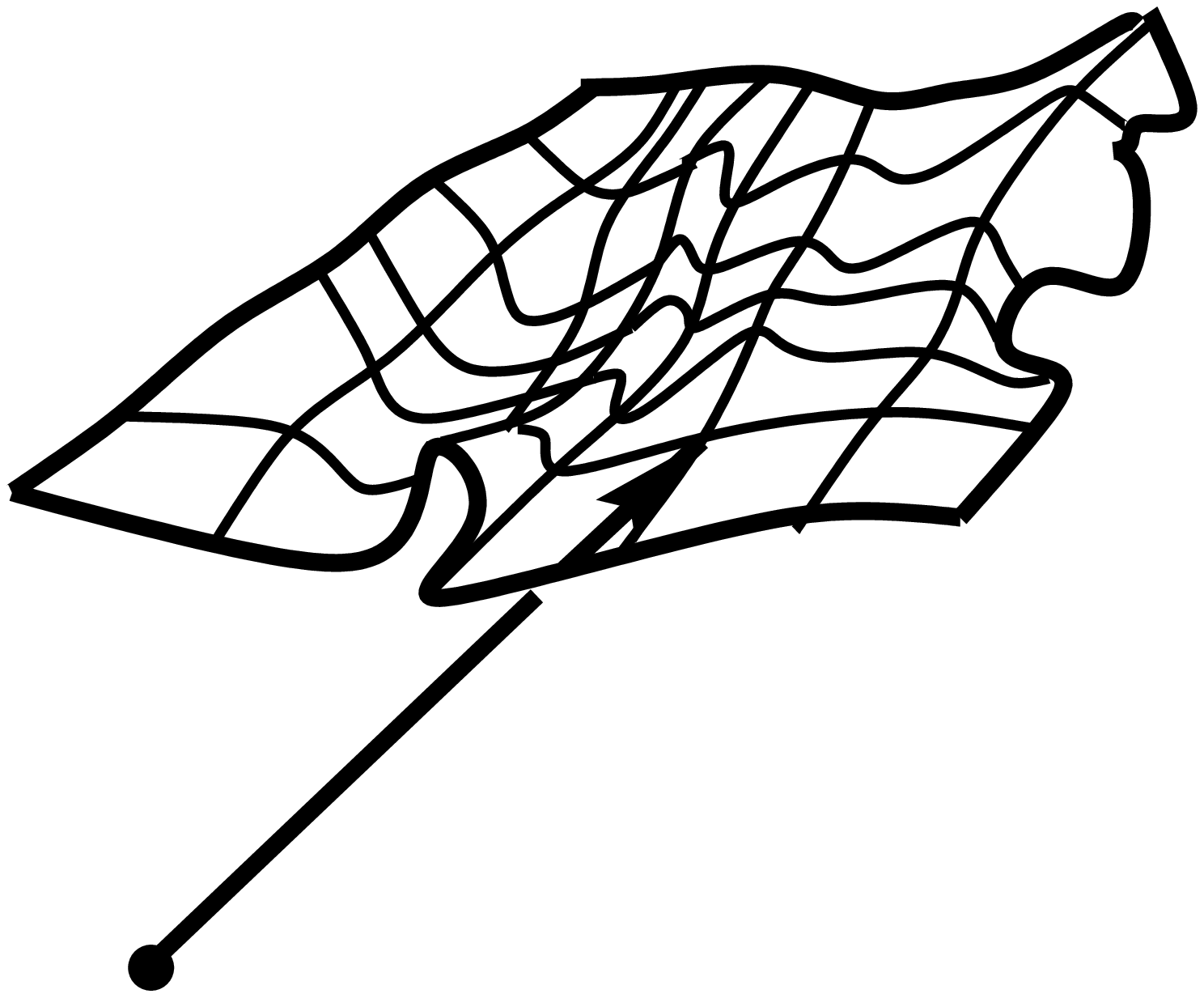}}
    \begin{picture}(0,0)
      \setlength{\unitlength}{1cm}
    \put(-5.2,0.0){{\small $0$}}
    \put(-4,1.2){{\small $\vec r(x)$}}
    \put(-2.56,2.56){\small $x$}
    \put(-2,0){(a)}
      \end{picture}
\end{center}
\end{minipage}
\hfill
\begin{minipage}[t]{8cm}
  \begin{center}
    \raisebox{0mm}{\includegraphics[scale=0.4]{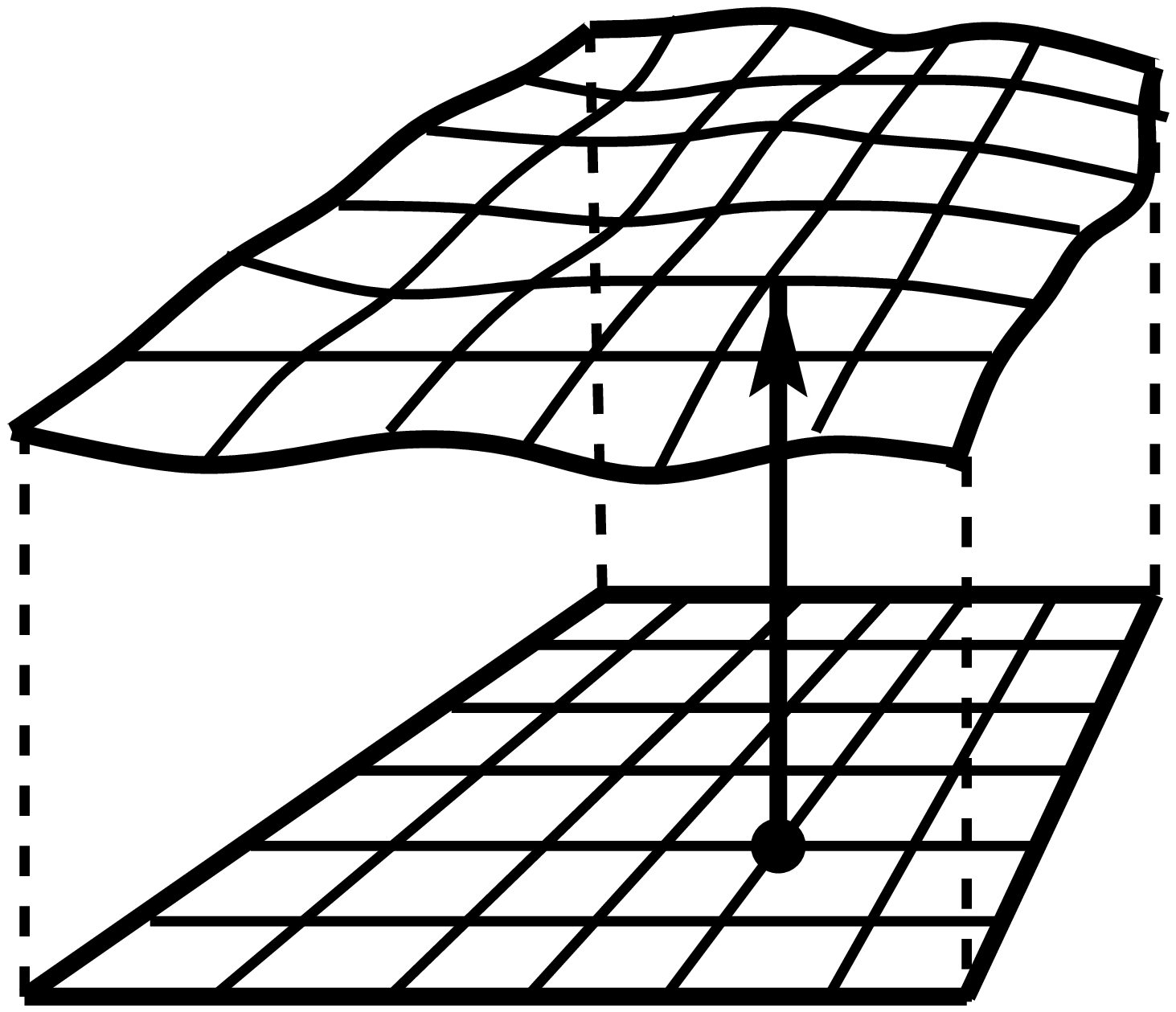}}
    \begin{picture}(0,0)
      \setlength{\unitlength}{1cm}
    \put(0,0){(b)}
    \put(-2.08,2.32){{\small $\vec r(x)$}}
    \put(-2,0.96){\small $x$}
    \put(-2.0,3.84){\small $x$}
      \end{picture}
\end{center}
\end{minipage}
   \caption{Left (a): A $D$-dimensional manifold ($D=2$) interacting with a point in the
     origin of the configurational space $\mathbb{R}^{d}$. Right (b): A
     ``directed'' membrane (interface) interacting with a parallel subspace of
     same dimension $D$.}
   \label{f2.00}
\end{figure}

\begin{figure}[t]
  \begin{center}
    \leavevmode
    \raisebox{0mm}{\includegraphics[scale=0.45]{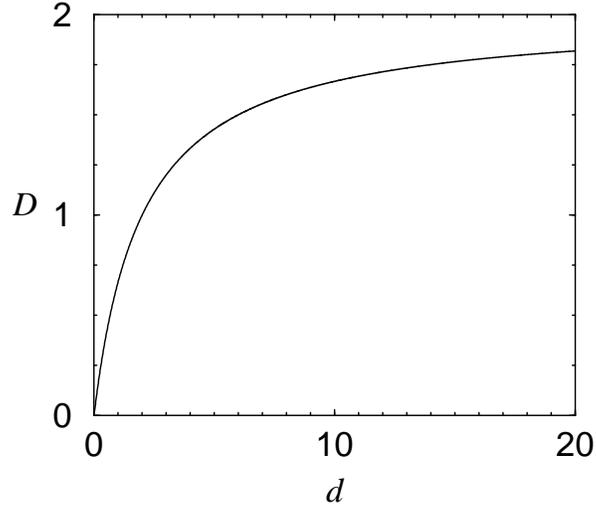}}  
\end{center}
    \caption{Critical line defined through $\varepsilon=0\Leftrightarrow
    d_{c}(D)=\frac{2D}{2{-}D}$. The interaction is relevant for points
that lie above that line.}
    \label{f2.0}
\end{figure}

The problem of a membrane avoiding
only a single point (\ref{I.1}) may at first sight appear
artificial. Besides of being a toy-model for the analysis of the more
interesting case of full self-avoidance as discussed above, let us
point out that it is, too, a special case of another more general
problem, which is interesting by its own. Consider a phantom tethered
membrane interacting with some $\delta$-potential located at the
origin of the configuration space: The Hamiltonian is given by ($0\le
D\le2$)
\begin{eqnarray}
  \label{2.0}
  \mathcal{H}[\vec r]=\frac{1}{2}\int\limits_{x\in\mathcal{M}}\left(\nabla
  \vec
  r(x)\right)^{2}+g_{0}\int\limits_{x\in\mathcal{M}}\delta^{d}(\vec
  r(x))\ ,
\end{eqnarray}
where any point in the membrane is labeled by some $D$-component vector $x$,
and its position in external space is given by the $d$-component field $\vec
r(x)$,
\begin{equation}
  \label{2.00}
  \vec r: x\in \mathbb{R}^{D}\longrightarrow \vec r(x)\in \mathbb{R}^{d}\ .
\end{equation}
The partition function is defined as
\begin{equation}
  \label{2.01}
{\cal  Z}  = \int {\cal D}\left[\vec r \right] \, \exp (-{\cal H}\left[\vec r
\right])\ .
\end{equation}
To remove the translational 0-mode, we will consider
\begin{equation}
  \label{2.02}
  {\cal Z}^{\diamond} = \int {\cal D}\left[\vec r \right] \ \delta(\vec r(y))\
  \exp (-{\cal H}\left[\vec r
\right])\ .
\end{equation}
We now discuss (\ref{2.0}) in more detail: The first term is the
elastic energy of the manifold which is entropic in origin. Elasticity
and temperature have been scaled to unity.  The second term models the
interaction of the manifold with a single point at the origin in the
$d$-dimensional configurational space. Let us remind
\cite{DDG1,PinnowWiese2001} that the physical interpretation depends
on the dimensionality: In the case that $\mathbb{R}^{d}$ is identical
to the embedding space, (\ref{2.0}) describes a phantom crumpled
manifold interacting with a single defect as sketched in figure
\ref{f2.00}. However, setting $d=1$ (\ref{2.0}) may as well describe a
solid-on-solid like fluctuating interface parameterized by some
displacement field and interacting with a parallel plane
($D=2$) as shown in figure \ref{f2.00}.\\
\vfill
 The coupling constant $g_{0}$ may either be positive (repulsive
interaction) or negative (attractive interaction).
We now give the dimensional analysis. 
In internal space units, the engineering dimensions are
\begin{eqnarray*}
  &&\mbox{dim}[x]=1  \\
  \nu&:=&\mbox{dim}[\vec r]=\frac{2-D}{2}
\end{eqnarray*}
\begin{equation}
  \label{2.2}
\varepsilon :=\mbox{dim}\left[ \int_{{\mathcal{M}}}\mbox{d}^{D}\!
  x\ \delta^{d}(\vec r(x))\right] =D-\nu d
\ .
\end{equation}
The interaction is naively
relevant for $\varepsilon>0$, i.e. $d<d_{c}$ with (see figure \ref{f2.0})
\begin{equation}
  \label{2.3}
  d_{c}=\frac{2D}{2{-}D}\ ,
\end{equation}
irrelevant for $\varepsilon<0$ and marginal for $\varepsilon=0$.
It has been  shown \cite{DDG1,DDG2} that the model is
renormalizable for $0<D<2$ and $\varepsilon\ge0$. Results for negative $\varepsilon$ 
are obtained via analytical continuation. 
\begin{figure}[b]
  \begin{center}
    \leavevmode
    \raisebox{0mm}{\includegraphics[scale=0.35]{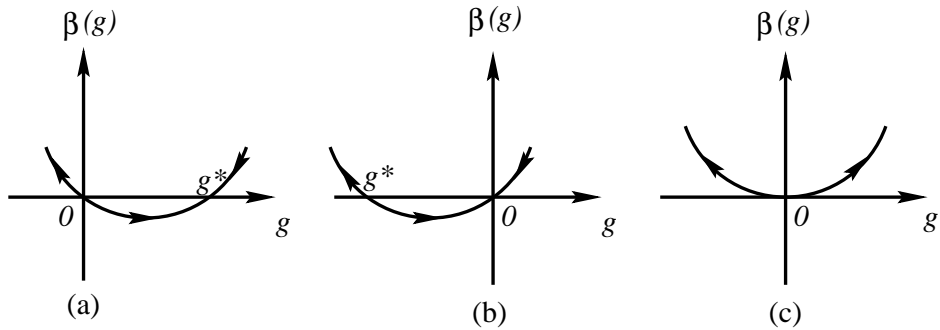}}
  \end{center} 
    \caption{RG-function and flow for increasing manifold size $L$ for
the dimensionless renormalized coupling $g$: (a) in the case
$\varepsilon{>}0$, (b) in the case $\varepsilon{<}0$, (c) in the case
$\varepsilon=0$. }
    \label{f2.1}
\end{figure}
One can define the renormalized coupling $g$ as
\begin{equation}
  \label{2.3.1}
  g:=\frac{\cal N}{\mathcal{V}_{\mathcal{M}}}\left[\mathcal{Z} (0)
-\mathcal{Z} (g_{0})\right]L^{\varepsilon}\ , 
\end{equation}
where $\mathcal{V}_{\mathcal{M}}$ denotes the internal volume of the
manifold. The normalization $\cal N$ depends on the definition of the
path-integral (but not on $L$) and is chosen such that
\begin{equation}\label{2.11p}
g = g_{0}L^{\varepsilon}+ O (g_0^{2})\ . 
\end{equation} 
Universal quantities emerge
at fixed-points of the $\beta$-function, which is defined as 
\begin{equation}
  \label{2.4}
  \beta(g):=-L \left.\frac{\partial g}{\partial L}\right|_{g_0}\ .
\end{equation}
The $\beta$-function thus  describes, how the effective coupling $g$
changes under scale 
transformations, while  keeping the bare coupling $g_{0}$ fixed. 
Let us give the 1-loop result, see e.g.~\cite{DDG1,DDG2,PinnowWiese2001}. 
It reads
\begin{equation}
  \label{2.5}
  \beta(g)=-\varepsilon g+\frac{1}{2}g^{2}+O(g^{3})\ ,
\end{equation}
where $g$ is the dimensionless renormalized coupling. Apart from the
trivial solution, $g=0$, the flow equation given by (\ref{2.4}) and
(\ref{2.5}) has a  non-trivial fixed point at the zero of the
$\beta$-function
\begin{equation}
  \label{2.6}
  g^{*}=2\varepsilon+O(\varepsilon^{2})\ .
\end{equation}
We shall show below that the scaling behavior is
described by the  slope of the RG-function at the fixed
point, which is universal as a consequence of renormalizabilty.
The long-distance behavior is then governed by the
$\delta$-interaction as considered in our model (\ref{2.0}), which is
the most relevant operator at large scales.  
Let us now discuss  possible physical situations (see fig.\ \ref{f2.1}):
\begin{itemize}
\item [(a)] $\varepsilon{>}0$: The RG-flow has an infrared stable
fixed point at $g^{*}>0$ and an IR-unstable fixed point at $g=0$. The
latter describes an unbinding transition whose critical properties are
given by the non-interacting system, while the non-trivial IR stable
fixed point determines the long-distance properties of the delocalized
state, the long-range repulsive force exerted by the fluctuating
manifold on the origin -- which we remind may be a point, a line or a
plane.
\item [(b)] $\varepsilon{<}0$: Now, the long-distance behavior is
Gaussian, while the unbinding transition occurs at some finite value
of the attractive potential, $g^{*}{<}0$, which corresponds to an
infrared unstable fixed point of the $\beta $-function.  Below $g^{*}$
the RG-flow is to strong coupling and the manifold is always
attracted.
\item [(c)] $\varepsilon=0$: This is the marginal situation, where the
transition takes place at $g^{*}=0$; we expect logarithmic corrections
to scaling.
\end{itemize}
These scenarios and possible observables have already been discussed
in \cite{PinnowWiese2001}. Here we want to specialize to the case of a
membrane avoiding a single point. It turns out that this situation
allows to calculate observables staying non-singular even for $2D$
membranes and which are accessible to a Monte-Carlo experiment.

\subsection{Plaquettes-density correlation
functions}\label{plaquettes-density}
Interesting physical observables for a membrane avoiding a
single point are the plaquettes-density functions at the repelling
point. Generally, these are defined as follows:
\begin{equation}
  \label{2.7}
  \left<n^{\ell}\right>_{\diamond}:=\left<\prod_{i=1}^{\ell}\ \int\limits_{x_{i}\in{\mathcal
  M}}\delta^{d}(\vec r(x_{i}))\right>_{\!\!\!\diamond}\ ,
\end{equation}
where the expectation value $\left<\cdot \right>_{\diamond}$ is taken
within the pinned ensemble as defined in (\ref{2.02}). The quantity,
which is  accessible in perturbation theory, is the
effective coupling as defined in (\ref{2.3.1}). It can be considered
as a generating function for observables like (\ref{2.7}). Let us
first show how to obtain the constrained partition function
(\ref{2.02}) from (\ref{2.3.1}): Since we consider closed manifolds,
internal translational invariance implies
\begin{equation}
  \label{2.8}
  \mathcal{Z}^{\diamond}\equiv{\mathcal Z}^{\diamond}(g_{0})=- \left. {}\frac{1}{{\mathcal V}_{{\mathcal M}}}
\frac{\partial}{\partial
  g_{0}}\right|_{L}\mathcal{Z}(g_{0})=
\frac{1}{\mathcal{N}} \left.\frac{\partial
(g L^{-\varepsilon})}{\partial 
  g_{0}}\right|_{L}\ ,  
\end{equation}
where $g$ is the renormalized or effective coupling defined in
(\ref{2.3.1}) and $\mathcal{V}_{\mathcal{M}}$ denotes the internal volume of
the membrane. Introducing the dimensionless bare coupling,
\begin{equation}
  \label{2.9}
  z:=g_{0}L^{\varepsilon}\ ,
\end{equation}
(\ref{2.8}) can be written in terms of dimensionless quantities as
\begin{equation}
  \label{2.10}
  {\mathcal Z}^{\diamond}(g_{0}L^{\varepsilon})=\frac{\partial g}{\partial z}\ ,
\end{equation}
where $\mathcal{N}$ has been set to unity. In the same way, all observables of the
type (\ref{2.7}) can be easily derived from $g$ according to:
\begin{equation}
  \label{2.11}
  \left<n^{\ell}\right>_{\diamond}=\frac{L^{\ell\varepsilon}}{\frac{\partial
  g}{\partial z}}\frac{\partial^{\ell}}{\partial{z^{\ell}}}\left(\frac{\partial
  g}{\partial z}\right)\ .
\end{equation}
An observable, which is accessible through
Monte-Carlo simulations, should be expressable by universal quantities like the
slope of the renormalization $\beta$-function (2.4). All these quantities can
be obtained from appropriate derivatives of the effective coupling $g$ with
respect to $z$: The $\beta$-function can be written in terms of the
bare coupling as
\begin{equation}
  \label{2.12}
  \beta(z)=-\varepsilon z\frac{\partial g}{\partial z}\ .
\end{equation}
(Note that in a slight abuse of notation, we write $\beta (z)=\beta (g
(z))$.)
The universal slope at the fixed point, which is defined as
\begin{equation}
  \label{2.13}
  \omega:=\left.\frac{\partial \beta(g)}{\partial g}\right|_{g^{*}}\ ,
\end{equation}
is obtained from
\begin{equation}
  \label{2.14}
  \omega(z)=\frac{-\varepsilon z}{\beta(z)}\frac{\partial \beta(z)}{\partial z}
\end{equation}
in the limit $z\to \infty$. We furthermore need the second derivative of the
RG-flow function with respect to the effective coupling, which is defined as
\begin{equation}
  \label{2.15}
  \omega^{\prime}:=\left.\frac{\partial^{2} \beta(g)}{\partial
  g^{2}}\right|_{g^{*}}\overset{z\to \infty}{=}\frac{-\varepsilon
  z}{\beta(z)}\frac{\partial \omega(z)}{\partial z}\ . 
\end{equation}
Let us now show that the universal slope (\ref{2.13}) is accessible through
the measurement of certain combinations of observables of type (\ref{2.7}). For
this purpose we need the plaquettes-density ($\ell=1$) and the density-density
function ($\ell=2$), which are obtained after some straight forward, but
tedious algebra from the above definitions:
\begin{eqnarray}
  \label{2.16}
  \left<n\right>_{\diamond}&=&\frac{1}{g_{0}}\left(1+\frac{\omega(z)}{\varepsilon}\right)\overset{z\to
  \infty}{\longrightarrow}\frac{1}{g_{0}}\left(1+\frac{\omega}{\varepsilon}\right)\nn\\
  \left<n^{2}\right>_{\diamond}&=&\frac{1}{g_{0}^{2}}\left(2+\frac{3\omega(z)}{\varepsilon}+\frac{\omega^{2}(z)}{\varepsilon^{2}}+\frac{\omega^{\prime}(z)\beta(z)}{\varepsilon^{2}}\right)\overset{z\to\infty}{\longrightarrow}\frac{1}{g_{0}^{2}}\left(2+\frac{3\omega}{\varepsilon}+\frac{\omega^{2}}{\varepsilon^{2}}\right)\ .
\end{eqnarray}
These quantities depend on the bare coupling $g_{0}$, which is not accessible. Instead, consider the following ratio:
\begin{equation}
  \label{2.17}
  \frac{\left<n\right>_{\diamond}}{\sqrt{\left<n^{2}\right>_{\diamond}}}\overset{z\to
  \infty}{=}\sqrt{\frac{1+\frac{\omega}{\varepsilon}}{2+\frac{\omega}{\varepsilon}}}\ ,
\end{equation}
which obviously is universal.

\subsection{Delocalization transition}\label{Delocalization
transition} For completeness let us shortly discuss the physical
situation at the UV-stable fixed point in figure \ref{f2.1}. The fixed
point corresponds to a {\em delocalization transition} of the
manifold, which is at vanishing coupling $g^{*}{=}0$ for
$\varepsilon{>}0$ and at some finite attractive coupling $g^{*}{<}0$
for $\varepsilon{<}0$.\\
In the localized phase $g{<}g^{*}$, correlation functions such as
$\left<[\vec r(x)-\vec r(y)]^{2}\right>$ and the associated
correlation length $\xi_{\parallel}$ (in the $D$-dimensional internal
space) should be finite, as well as the radius of gyration
$\xi_{\perp}$. Approaching the transition these quantities diverge as
\cite{ForgasLipowskyNieuwenhuizenInDombGreen}
\begin{equation}
  \label{u.1}
  \xi_{\parallel}\ \sim \ (g^{*}{-}g)^{-\nu_{\parallel}}\quad ,\quad
  \xi_{\perp}\ \sim \ (g^{*}{-}g)^{-\nu_{\perp}}\ .
\end{equation}
Since $\xi_\perp\sim\xi_\parallel^{\nu }$, the exponents $\nu_{\parallel}$ and $\nu_{\perp}$ are related through
\begin{equation}
  \label{u.2}
  \nu_{\perp}\ =\ \nu_{\parallel}\nu\ ,
\end{equation}
$\nu$ being the dimension of the field (\ref{2.2}).

Furthermore, they are related to the correction-to-scaling exponent $\omega$: 
\begin{equation}
\nu _{\parallel } = -\frac{1}{\displaystyle \omega (g^{*})} \ , \qquad
\nu _{\perp  } = -\frac{\nu }{\displaystyle \omega (g^{*})} \ .
\end{equation}
Note that $\omega(g^{*}){<}0$ at the transition. Specializing to
$(D,d){=}(1,1)$, we find
\begin{equation}
  \label{u.6}
  \nu_{\perp}\ =\ 1\quad ,\quad \nu_{\parallel}\ = \ 2\ .
\end{equation}

These exponents are also valid for  the delocalization transition of a
$1$-dimensional interface from an attractive hard wall in $2$-dimensional bulk
space \cite{BrezinHalperinLeibler1983,ForgasLipowskyNieuwenhuizenInDombGreen,Upton1999,PinnowWiese2001}. 


\section{Complete summation of the perturbation series}\label{complete resummation}
\subsection{Perturbation theory}\label{perturbation theory}

In (\ref{plaquettes-density}) we saw that physical observables can be derived
from the renormalized coupling $g$ (\ref{2.3.1}). To obtain $g$ we need the
perturbation series of the partition function ${\mathcal Z}$ (\ref{2.01}):
\begin{equation}
  \label{3.3}
  \mathcal{Z}=\sum_{N=-1}^{\infty}\frac{(-g_{0})^{N{+}1}}{(N{+}1)!}\mathcal{Z}_{N}\ ,
\end{equation}
where
\begin{equation}
  \label{3.4}
  \mathcal{Z}_{N}=\left<{\prod^{N{+}1}_{i=1}\int_{x_{i}}\
  \tilde\delta^{d}(r(x_{i}))}\right>_{\!\!\!0}\ , \qquad N\geq 0\ ,
\end{equation}
and the normalization of the $\delta$-distribution has been chosen to be
\begin{equation}
  \label{3.4.1}
  \tilde \delta^{d}(r(x)):=(4\pi)^{d/2}\delta(r(x))=\int_{k}\rme^{ikr(x)}
\end{equation}
with
\begin{equation}
  \label{3.4.2}
  \int_{k}:=\pi^{-d/2}\int \mbox{d}^{d}k
\ .
\end{equation}
The  advantage of these normalizations is that 
\begin{equation}
\int_{k} \rme^{-k^{2}} = 1\ .
\end{equation}
Accordingly, the perturbation expansion of the effective coupling (\ref{2.3.1}) reads
\begin{equation}
  \label{3.4.3}
g(z)=\frac{g_{0}L^{\varepsilon}}{\mathcal{V_{\mathcal{M}}}}
\sum_{N=0}^{\infty}\frac{(-g_{0})^{N}}{(N{+}1)!}
\left<\prod_{i=1}^{N{+}1}\int_{x_{i}}\tilde\delta^{d}(r(x_{i}))\right>_{0}
\ .
\end{equation}
Performing the averages within the Gaussian theory with normalization
\begin{equation}
  \label{3.4.4}
  \frac{1}{{\cal V}_{\cal
M}}\int_{x} \left< \tilde\delta^{d}\left( r (x) \right)\right>_{0}=1\ ,
\end{equation}
one arrives at
\begin{equation}
  \label{3.4.5}
g(z)=\frac{g_{0}L^{\varepsilon}}{\mathcal{V_{\mathcal{M}}}}
\sum_{N=0}^{\infty}\frac{(-g_{0})^{N}}{(N{+}1)!}\left(\prod_{i=1}^{N{+}1}\
\int\limits_{\!  k_{i}}\!\!\int\limits_{x_{i}} \right)
\tilde\delta^{d}\big(\sum_{i=1}^{N{+}1}k_{i}\big)\
\mathrm{e}^{\frac{1}{2}\!\!\!\sum
\limits_{i,j=1}^{N{+}1}k_{i}k_{j}C(x_{i}{-}x_{j})}\ ,
\end{equation}
where
\begin{equation}
  \label{3.4.6}
  C(x_{i}{-}x_{j}):=\frac{1}{2d}\left<( r(x_{i})-r(x_{j}))^{2}\right>_{0}
\end{equation}
denotes the  correlator, and the $\tilde\delta^{d}(\sum_{i}k_{i})$
stems from the integration over the global translation. Shifting
\begin{equation}
  k_{N{+}1}\to k_{N{+}1}-\sum_{i=1}^{N}k_{i}\ ,
\end{equation}
the quadratic form in (\ref{3.4.5}) transforms to
\begin{eqnarray}
  \frac{1}{2}\sum_{i,j=1}^{N{+}1}&&k_{i}k_{j}C(x_{i}{-}x_{j})\nonumber \\
&&{\longrightarrow}\sum_{j=1}^{N}k_{N{+}1}k_{j}C(x_{N{+}1}{-}x_{j}){-}\sum_{i,j=1}^{N}k_{i}k_{j}\frac{C(x_{N{+}1}{-}x_{i}){+}C(x_{N{+}1}{-}x_{j}){-}C(x_{i}{-}x_{j})}{2}\ .\nn\\
\label{3.4.7}
\end{eqnarray}
Integrating out the momenta $k_{1},\dots,k_{N{+}1}$ in (\ref{3.4.5}),
one obtains
\begin{equation}
  \label{3.4.8} g(z)=z\sum_{N=0}^{\infty}\frac{(-z)^{N}}{(N{+}1)!}
\left(\prod_{\ell=1}^{N}\ \int\limits_{x_{\ell}} \right)(\det
\mathfrak{D})^{-d/2}\ ,
\end{equation}
where we have factored out $L^{\varepsilon}$ from the loop integration
 (such that the integrals now run over a torus of size 1), and
the matrix elements $\mathfrak{D}_{ij}$ are
\begin{eqnarray}
  \mathfrak{D}_{ij}=\frac{1}{2}[C(x_{N{+}1}{-}x_{i}){+}C(x_{N{+}1}
    {-}x_{j}){-}C(x_{i}{-}x_{j})]\ .
\end{eqnarray}

\subsection{Complete summation in fixed internal space dimension
$D=2$}\label{direct} Let us compute the $N$-loop order of
(\ref{3.4.8}): The behavior of the propagator $C(x)$ for arguments $x$
large compared to $a$ is of the form 
\begin{equation}
  C(x) =  c_{0}+\frac{1}{2\pi}\ln\frac{x}{a} \ ,
\end{equation}
where $c_{0}$ denotes some positive constant (note $C (x)\ge0$), and
the logarithmic growth (for large $x$) is universal (see appendix
\ref{sec:ker}).  In $D=2$ we need an additional short-distance cutoff
$a$, which we want to take to 0. 
 We can (somehow arbitrary) decompose
\begin{equation}
 \det \mathfrak{D}=(\prod_{i=1}^{N}\mathfrak{D}_{ii})\det
\mathfrak{\tilde D} \ .
\end{equation}
In the limit of $a \to 0$ each $C (x) = \frac{1}{2\pi}\ln (L/a) +O
(a^{0})$, such that
\begin{eqnarray}\label{com.0.1} \mathfrak{\tilde
D}_{ij}&=&\frac{1}{2}\left[1{+}\frac{C(x_{N{+}1}{-}x_{j})-C(x_{i}{-}x_{j})}
{C(x_{N{+}1}{-}x_{i})}\right]\xrightarrow{a\to
0}\frac{1}{2}\ ,\
    i{\not=}j\ , \nonumber\\
  \mathfrak{\tilde D}_{ii}&=&1\ .
\end{eqnarray}
\begin{equation}\label{com.1} 
\left(\prod_{\ell=1}^{N}\,\int\limits_{x_{\ell}} \right) (\det
\mathfrak{D})^{-d/2}=: I_{N}(L/a)=I_{1}^{N}(L/a)\
(\det\mathfrak{\tilde D}^{(0)})^{-d/2} \ .
\end{equation}
The matrix $ \mathfrak{\tilde D}^{(0)}$ denotes the limit $a\to 0$ of
(\ref{com.0.1}). It can be written as $\mathfrak{\tilde
D}^{(0)}=\frac{1}{2}(\mathbb{I}+N\mathbb{P})$, where $\mathbb{I}$
denotes the identity and $\mathbb{P}$ the projector onto
$(1,1,\dots,1)$, whose image has dimension $1$, such that $\det
\mathfrak{\tilde D}^{(0)}=\frac{1{+}N}{2^{N}}$ \cite{PinnowWiese2001}. Furthermore, to one
loop $I_{1}(L/a)\overset{a\to 0}{=}c_{1}(\ln\frac{L}{a})^{-d/2}$,
where $c_{1}$ denotes some (finite) constant. One then arrives at
\begin{equation}
  \label{com.2}
g(z)=z\sum_{N=0}^{\infty}\frac{(-z(\ln\frac{L}{a})^{-d/2})^{N}}
{N!(1{+}N)^{d/2{+}1}}
\ .
\end{equation}
A factor $c_{1}2^{d/2}$ has 
been absorbed into a rescaling of both $z$ and $g$. 

\subsection{Asymptotic scaling behavior}
In the following we will analyze the limit of large $z$
(strong repulsion), which also is the  scaling behavior of infinitely
large membranes. We  need an analytical expression for sums like
(\ref{com.2}) in the limit of large $z$. Later, it will turn out that allowing
for small deviations $2-D>0$ only slightly more general sums will
arise.

\begin{figure}[t]
  \begin{center}
    \vspace{1cm}
    \setlength{\unitlength}{1cm}
    \begin{picture}(7,5)
    \put(0,0){\raisebox{0mm}{\includegraphics[scale=0.8]{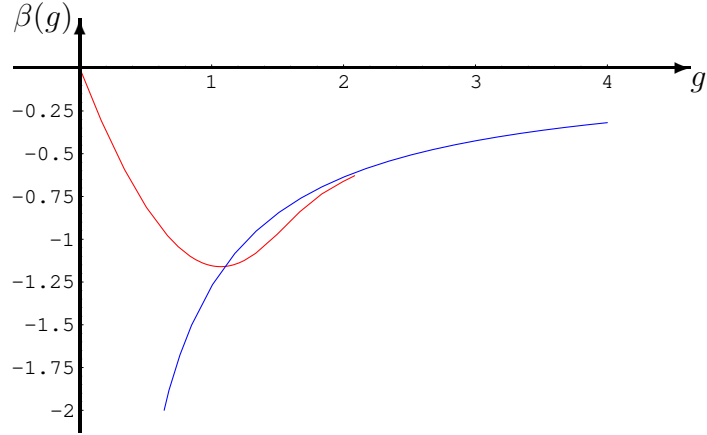}}}
    \thicklines
    \put(0,4.65){\vector(1,0){9.0}}
    \put(0.88,-0.2){\vector(0,1){5.5}}
    \put(9.0,4.4){$g$}
    \put(-0.0,5.2){$\beta(g)$}
      \end{picture}
    \end{center} 
\caption{$\beta-$function in terms of the
    renormalized coupling $g$  truncated at order 160, Pade-resummed,
and plotting 
only that part for which the truncated series converges. (This can
e.g.\ be tested by taking away the last few terms of the series.) This
is  compared to the  asymptotic behavior (\ref{as.4})
    (proportional to $1/ g$ for large $g$).  $d$ is set to $1$,
and we used the diagonal (80,80)-Pade approximant, which was find to
converge best. (The non-resummed expression starts to diverge already at
$g\approx 1.8$ at this order.)} 
    \label{betapa}
\end{figure}

We claim that for all $k,d>0$
\begin{equation}\label{6.28}
  \sum_{N=0}^{\infty}\frac{(-z)^{N}}{N!(k+N)^{d/2}}=\frac{1}
{\Gamma(\frac{d}{2})}\int_{0}^{\infty}\mbox{d}r\ r^{d/2-1}\exp[-z\
\rme^{-r }-k r]\ .
\end{equation}
This can be proven as follows:
\begin{eqnarray*}
  \frac{1}{\Gamma(\frac{d}{2})}\int_{0}^{\infty}\mbox{d}r\
r^{d/2-1}\exp ( -z\, \rme^{-r}-k
r)&=&\frac{1}{\Gamma(\frac{d}{2})}\sum_{N=0}^{\infty}\frac{(-z)^{N}}
{N!}\int_{0}^{\infty}\mbox{d}r\   r^{d/2-1}\rme^{-(N+k)r}\\
  &=&\frac{1}{\Gamma(\frac{d}{2})}\sum_{N=0}^{\infty}\frac{(-z)^{N}}{N!}
\frac{\Gamma(\frac{d}{2})}{(N+k)^{d/2}}\ .
\end{eqnarray*}
This integral-representation is not the most practical for our
purpose. It is better to set $r\to s:=\rme^{-r}$ which yields 
\begin{equation}
  \sum_{N=0}^{\infty}\frac{(-z)^{N}}{N!(k+N)^{d/2}}=\frac{1}{\Gamma
(\frac{d}{2})}\int_{0}^{1}\mbox{d}s\, s^{k-1} (-\ln s)^{d/2-1}
\rme^{-s z}\ .
\end{equation}
This formula is already very useful for some purposes. It is still
advantageous to make a second variable-transformation $s\to y := sz$, yielding
\begin{equation}
  \sum_{N=0}^{\infty}\frac{(-z)^{N}}{N!(k+N)^{d/2}}=\frac{(\ln z)^{d/2-1}}{\Gamma(\frac{d}{2})z^{k}}\int_{0}^{z}\mbox{d}y\, y^{k-1} \left(1-\frac{\ln y}{\ln z} \right)^{d/2-1} \rme^{-y}\ .
\end{equation}
Finally we remark that we usually have the following combination
\begin{equation}
f_{k}^{d} (z):= z^{k}
\sum_{N=0}^{\infty}\frac{(-z)^{N}}{N!(k+N)^{d/2}}=\frac{(\ln
z)^{d/2-1}}{\Gamma(\frac{d}{2})}\int_{0}^{z}\rmd y\, y^{k-1}
\left(1-\frac{\ln y}{\ln z} \right)^{d/2-1} \rme^{-y}\ . 
\end{equation}
It satisfies the following simple recursion relation, which is
helpful to calculate the $\beta$-function:
\begin{equation}\label{recursion}
z \frac{\rmd }{\rmd z} f_{k}^{d} (z) = f_{k}^{d-2} (z) \ . 
\end{equation}
The derivative above can also be rewritten as
\begin{equation}
  \label{recursion2}
  z \frac{\rmd }{\rmd z} f_{k}^{d} (z) = kf_{k}^{d} (z)-f_{k{+}1}^{d} (z)\ ,
\end{equation}
such that one obtains a useful formula in order to isolate the dominant
behavior for large $z$:
\begin{equation}
  \label{fkdreduce}
  f_{k{+}1}^{d} (z) = kf_{k}^{d} (z)-f_{k}^{d{-}2} (z) \ .
\end{equation}
From (\ref{6.28}) $f_k^{d} (z)>0$ for all $k,d>0$ and the behavior for
large $z$ is obtained by expanding $\left(1-\frac{\ln y}{\ln z}
\right)^{d/2-1}$ for small $\frac{1}{\ln z}$
\begin{eqnarray}
  \label{fkdexpansion}
f_{k}^{d} (z) &=\ds \frac{(\ln z)^{d/2-1}}{\Gamma(\frac{d}{2})} \Bigg[&
\int_{0}^{\infty } \rmd y\, y^{k-1} \rme^{-y} \nonumber \\
&& -\frac{1}{\ln z}\left(\frac{d}{2}-1 \right)\int_{0}^{\infty }
\rmd y\, y^{k-1}\ln y\, \rme^{-y}\nonumber \\
&& +\, O  \left(\frac{1}{(\ln z)^{2}} \right)
 \Bigg] +\, O (\rme^{-z})\ .
\end{eqnarray}
The result is 
\begin{equation}
  \label{fkdasympt}
f_{k}^{d} (z) =\frac{(\ln z)^{d/2-1}\Gamma (k)}{\Gamma(\frac{d}{2})}
\left(1- \frac{1}{\ln z} \frac{d{-}2}{2} \frac{\Gamma' (k)}{\Gamma
(k)} + \dots  \right)
\ .
\end{equation}
With the above notations, the sum (\ref{com.2})
expressing $g$ as a function of $z$ becomes 
\begin{eqnarray}\label{star}
g \left(z, L/a\right) &=& \left(\ln \frac{L}{a}\right)^{d/2}\
f_{1}^{{d+2}} \left[z\ \left(\ln \frac{L}{a}\right)^{-d/2}\right]\ .
\end{eqnarray}
in the limit $D=2$.\\
It is now easy to analyze the long-distance behavior in this limit. First, we
observe that according to (\ref{fkdasympt}) the effective coupling diverges
logarithmically for all external dimensions $d>0$:
\begin{equation}
  \label{as.1}
  g\left(z, L/a\right)~\xrightarrow{z\to\infty}~\frac{\ \ \big(\!\ln{\textstyle
\frac{L}{a}}\big)^{d/2}}{\Gamma(\frac{d{+}2}{2})}\left[\ln\left(z\left(
\ln{\textstyle \frac{L}{a}}\right)^{-d/2}\right) \right]^{d/2}\ .
\end{equation}
This is in contrast to the one-loop result as stated in (\ref{2.6}),
which is exact for polymers ($D=1$) and which stays qualitatively
valid as long as $D<2$. This follows from the renormalizability of the
theory \cite{DDG1} for sufficiently small $\varepsilon>0$. A finite
limit $g(z\to \infty)=g^{*}$ signals a scale invariant theory. In
(\ref{as.1}) we have found the limiting behavior of the
latter. Consequently, we expect the correction-to-scaling exponent
$\omega$ to be always zero in $D=2$. In order to check that let us
first compute the renormalization $\beta$-function in terms of the
bare coupling as in (\ref{2.12}), which can be immediately derived
with the help of relation (\ref{recursion})\footnote{Note that our
definition $\beta (z')= - \epsilon z' \partial g'/\partial z'$ is strictly
speaking equivalent to definining the $\beta$-function as $\beta (g):=
\left.\left(-L \frac{\rmd}{\rmd L} - a \frac{\rmd}{\rmd a}
\right)\right|_{g_{0}} g$, instead of (\ref{2.4}). (Note that the
derivative w.r.t.\ $a$ disappears for $D<2$.) The natural combination
in $D=2$ is $z'=g_{0}L^{\epsilon } \left(\ln \frac{L}{a}
\right)^{-d/2}$ instead of $z=g_{0}L^{\epsilon }$, and normalizations
such that $g' (z') =z' + O (z'^{2})$ does not explicitly depend on $L$
or $a$. The chosen defintions avoid unnecessary technical
complications, but do not change the physics of the problem.}:
\begin{equation}
  \label{as.2}
  \beta(z')=-\varepsilon z'\frac{\partial g'}{\partial z'}= -\varepsilon\
  f_{1}^{d}(z')~\xrightarrow{z'\to\infty}~\frac{1}{\Gamma(\frac{d}{2})}\left[\ln (
  z')\right]^{d/2-1}\ ,
\end{equation}
where we have introduced rescaled couplings $z':=g_{0}L^{\varepsilon}(\ln \frac{L}{a})^{-d/2}$ and $g'=g\ (\ln \frac{L}{a})^{-d/2}$. Its derivative with respect to the renormalized coupling is found as a
function of the bare coupling (\ref{2.14}) to be
\begin{eqnarray}
  \label{as.3}
  &&\omega(z')=\frac{-\varepsilon z'}{\beta(z')}\frac{\partial \beta(z')}{\partial
    z'}=-\varepsilon \frac{z'\frac{\rmd}{\rmd
  z'}f_{1}^{d}(z')}{f_{1}^{d}(z')}~\xrightarrow{z'\to\infty}~\varepsilon \frac{2-d}{2\ \ln
  (z')} ~\xrightarrow{z'\to\infty}~ 0\ .
\end{eqnarray}
Note that the qualitative behavior of the $\beta$-function changes
depending on the external dimension $d$,
approaching asymptotically zero below $d=2$ and being divergent above.\\
In the limit of large bare couplings one may as well give the
RG-function in terms of the effective (renormalized) coupling simply
by inverting the asymptotic expression in (\ref{as.1}) and inserting
it into (\ref{as.2}), with the result:
\begin{equation}
  \label{as.4}
  \beta(g)\overset{z'\to
  \infty}{\sim}-\varepsilon\frac{(\Gamma(\frac{d{+}2}{2}))^{1-2/d}}{\Gamma(\frac{d}{2})}\
  g^{1-2/d}\ .
\end{equation}
It is interesting to compare the true asymptotic behavior of the
completely resummed perturbation series as found above with
predictions taking only finite loop orders into account: If one tries
to invert (\ref{com.2}) and truncates it at some finite order, it is
at least possible to reach the asymptotic regime (\ref{as.4}) --
however, for large $g$ the truncated $\beta-$function does not
converge to the true $\beta$-function and thus strongly deviates from
the true behavior. In figure \ref{betapa} the Pade-resummed truncated
$\beta$-function up to order $g^{160}$ in $d=1$ is compared with the
asymptotic flow-function. One notices that the truncated
$\beta$-function even though improved through a Pade-Resummation
hardly gets into touch with the asymptotic regime. The same applies to
the slope-function $\omega(g)$, which is not shown in
figure \ref{betapa}.
Let us finally state the expected behavior of the plaquettes-density
functions in the limit of large membranes. For the plaquettes-density
at the repelling fixed-point we find in this limit:
\begin{equation}
  \label{as.5}
  \left< n\right>_{\diamond}=\frac{1}{g_{0}}\left(1+\frac{2-d}{2\ \ln
  z}\right)\overset{z\to \infty}{\sim}\frac{1}{g_{0}}\ .
\end{equation}
Note that in the absence of the repelling interaction this quantity would
diverge in this limit. This follows from dimensional grounds, since then
\begin{equation}
  \label{as.6}
  \left< n\right>_{\diamond}\sim L^{\varepsilon}\ .
\end{equation}
In (\ref{as.5}) we found the largest possible depopulation  of monomers at the defect potential in  the case
of a relevant interaction ($\varepsilon>0$). As we discussed in
(\ref{plaquettes-density}) a measurable quantity should be the following
ratio (\ref{2.17}), which in the case of $2D$-membranes becomes in the limit
$z\to \infty$:
\begin{equation}
  \label{as.6.b}
  \frac{\left<n\right>_{\diamond}}{\sqrt{\left<n^{2}\right>_{\diamond}}}\overset{z\to
  \infty}{=}\sqrt{\frac{1}{2}}\ ,
\end{equation}
which can be compared with the 1-loop prediction (which is exact for
polymers):
\begin{equation}
  \label{as.7}
  \frac{\left<n\right>_{\diamond}}{\sqrt{\left<n^{2}\right>_{\diamond}}}\overset{z\to
  \infty}{=}\sqrt{\frac{2}{3}}\ , \quad \textrm{(1-loop)}\ .
\end{equation}

\section{Crossover to polymers}\label{2mDtorus}
\begin{figure}[t]
\centerline{\includegraphics[width=0.5\textwidth]{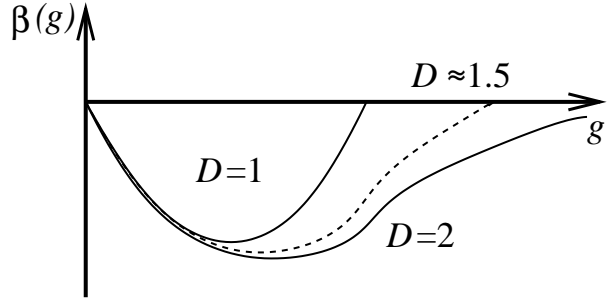}}
\caption{Qualitative behavior for the $\beta $-function in $D=1$, $D=2$
and result anticipated for $D\approx 1.5$.}
\label{guessbeta}
\end{figure}

 Let us now analyze the
theory below $D=2$. Due to the renormalizability in $0<D<2$ and the
existence of an $\varepsilon$-expansion we expect the renormalized
coupling to reach a finite fixed point in the strong coupling limit as
soon as $D<2$. This approach is characterized by a power-law decay of
the form
\begin{equation}\label{2mD.0}
  g(z)=g^{*}+S(\ln z)\ z^{-\omega/\varepsilon}+
O(z^{-\omega_{1}/\varepsilon})\ ,
\end{equation}
where $S$ is some scaling-function growing at most sub-exponentially
and $\omega_{1} >\omega >0$, with $\omega$ defined in (\ref{2.13}).
\\
Our ultimate aim is to draw information from an expansion in powers of
$2-D$ of the effective coupling about the correction-to-scaling
exponent $\omega$ in (\ref{2mD.0}) for $D\lesssim 2$. The scale
invariant behavior below $D=2$ results in a finite fixed point of the
renormalization $\beta$-function as a function of the effective
coupling. The qualitative behavior of the $\beta$-function is sketched
in fig. (\ref{guessbeta}).
 \\
\subsection{$(2{-}D)$-expansion on the torus}\label{s:D2} In order to
gain information about $g$ below $D=2$ one has to expand the loop
integrand $(\det \mathfrak{D})^{-d/2}$ (\ref{3.4.8}) in powers of
$2{-}D$. For convenience, we take $a\to 0$. The propagator takes in
infinite $D$-space the form $C(x)=|x|^{2{-}D}/(S_{D}(2{-}D))$, where
$S_{D}=2\pi^{D/2}/\Gamma(\frac{D}{2})$ denotes the volume of the
$D$-dimensional unit-sphere. The factor $(S_{D}(2{-}D))^{-1}$ replaces
$\ln (\frac{L}{a})$ and is absorbed into a rescaling of the field and
the coupling according to $r\to r\ (S_{D}\ (2{-}D))^{1/2}$ and
$g_{0}\to g_{0}\ (S_{D}\ (2{-}D))^{d/2}$, such that the factors of
$(\ln\frac{L}{a})^{-d/2}$ in (\ref{com.2}) and (\ref{as.1}) disappear.
The propagator in the rescaled
variable can then be written as
\begin{equation}
  \label{2mD.1}
  C(x)=1+(2-D)\ \mathbb{C}(x)
\ ,
\end{equation}
where for convenience of notation we allow $\mathbb{C} (x)$ to depend
itself on $D$. 
\\

Of course, on a closed manifold of finite size, $C(x)$ is modified,
but the form (\ref{2mD.1}) is independent of the shape of the
manifold. Accordingly, one may expand the matrix $\mathfrak{D}$ as
\begin{equation}
  \label{2mD.1.1}
 \mathfrak{D}=\mathfrak{\tilde D}^{(0)}+(2{-}D)\ \mathbb{D}
\ ,
\end{equation}
where
$\mathfrak{\tilde D}^{(0)}$ is defined as before and coincides with the
limit $D{\to}2$ when inserting the above $C(x)$ into
$\mathfrak{D}$. Moreover, $\mathbb{D}$ is of the same form as $\mathfrak{D}$,
but each $C (x)$ has been replaced with $\mathbb{C}(x)$:
\begin{equation}
  \label{2mD.1.2}
  \mathbb{D}_{ij}=\frac12 [\mathbb{C} (x_{N+1}-x_{i})+\mathbb{C}
(x_{N+1}-x_{j})+\mathbb{C} (x_{i}-x_{j}) ]\ .
\end{equation}
Then,
\begin{equation}\label{2mD.2}
\det \mathfrak{D}=\det 
\mathfrak{\tilde D}^{(0)}\, \exp\left\{ \mathrm{Tr}\left[  \ln(1+(2-D)[
\mathfrak{\tilde D}^{(0)}]^{-1}\mathbb{D}) \right] \right\}\ ,
\end{equation}
where $[ \mathfrak{\tilde
D}^{(0)}]^{-1}=2(\mathbb{I}{-}\frac{N}{N{+}1}\mathbb{P})$ denotes the
inverse matrix of $\mathfrak{\tilde D}^{(0)}$.
\\
Denoting
\begin{equation}
  \label{6.62}
  \mathfrak{M}:=[\mathfrak{D}^{(0)}]^{{-1}}\mathfrak{D}\ ,
\end{equation}
we expand the determinant in (\ref{2mD.2}) up to fourth order in
$2{-}D$: 
\begin{eqnarray}
        \label{6.71}
        &&[\det(\mathfrak{D})]^{-d/2}=[\det(\mathfrak{D}^{(0)})]^{-d/2}\left[1-\frac{d}{2}\left[(2-D)\mbox{Tr}\,\mathfrak{M}\right.\right.\nonumber\\
        &&\left.-\frac{(2-D)^2}{2}\mbox{Tr}\,\mathfrak{M}^2+\frac{(2-D)^3}{2}\mbox{Tr}\,\mathfrak{M}^3-\frac{(2-D)^4}{4}\mbox{Tr}\,\mathfrak{M}^4\right]\nonumber\\
        &&+\frac{d^2}{8}\left[(2-D)^2\mbox{Tr}^2\,\mathfrak{M}-(2-D)^3 \mbox{Tr}\,\mathfrak{M}\ \mbox{Tr}\,\mathfrak{M}^2+(2-D)^4\left[\frac{1}{4}\mbox{Tr}^2\,\mathfrak{M}^2+\frac{2}{3}\mbox{Tr}\,\mathfrak{M}\ \mbox{Tr}\,\mathfrak{M}^3\right]\right]\nonumber\\
        &&\left.-\frac{d^3}{48}\left[(2-D)^3\mbox{Tr}^3\,\mathfrak{M}-(2-D)^4
\frac{3}{2}\mbox{Tr}^2\,\mathfrak{M}\
        \mbox{Tr}\,\mathfrak{M}^2\right]+\frac{d^4}{384}(2-D)^4\mbox{Tr}^4\,\mathfrak{M}\right]+O((2-D)^5)\ . \nonumber\\
    &&
\end{eqnarray}
The first step in the analysis will be to obtain the resummed
perturbation series of the effective coupling up to fourth order in
$2-D$. That is, we have to insert (\ref{6.71}) into (\ref{2mD.2}),
calculate the corresponding loop integrals at each order of
perturbation theory, insert the result into (\ref{3.4.8}) and sum the
appearing series to all
orders.\\
\\
Let us start with the first-order term in $2-D$ from
(\ref{6.71}). We only need $ \mathfrak{M}=[\mathfrak{D}^{(0)}]^{-1} \mathfrak{D}$, which reads
\begin{eqnarray}
  \label{6.11}
 (\mathfrak{M}_{ij})=\left( [\mathfrak{D}^{(0)}]^{-1} \mathfrak{D}\right)_{ij}=\left(2\mathfrak{D}_{ij}
-\frac{2}{1+N}\sum_{k=1}^{N}\mathfrak{D}_{ik}\right)L^{-2\nu}\ . 
\end{eqnarray}
The trace of (\ref{6.11}) can easily be performed, with the result
\begin{eqnarray}
  \label{6.12} \mbox{Tr}
\,\mathfrak{M}=\left(\frac{2N}{1+N}\sum_{i=1}^{N}\mathfrak{D}_{ii}-
\frac{2}{1+N}\sum_{i=1}^{N}\sum_{k=1}^{N}(1-\delta_{ik})\mathfrak{D}_{ik}\right)
L^{-2\nu} \ .
\end{eqnarray}
In each order of perturbation theory we have to integrate the
expression (\ref{6.71}) over internal distances. These integrals have
to be regularized in the infrared through an appropriate IR cut-off.
We are considering a finite manifold of toroidal topology.  The
precise form of the correlator on the torus will only later enter into
the calculation.
\begin{figure}[t]
\begin{center}
\raisebox{-10mm}{\includegraphics[scale=0.4]{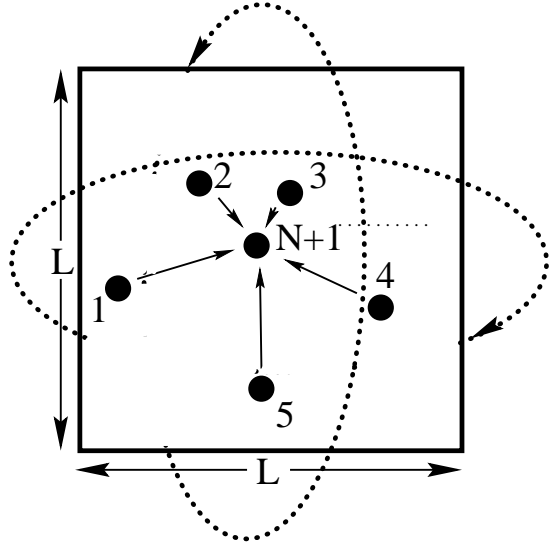}}
\end{center}
\caption{Regularization scheme for the $N$-loop diagrams on manifolds with
  toroidal topology (periodic boundary conditions). Here: $D=2$}
    \label{f6.1}
\end{figure}

To simplify the calculations, we further introduce the following notation:
\begin{equation}
  \label{6.13.1}
  \overline{f(x_{i_{1}},\dots,x_{i_{k}})} := \int_{x_{1}}\cdots\int_{x_{N}}f(x_{i_{1}},\dots,x_{i_{k}})
\end{equation}
with  the internal integrations defined as 
\begin{equation}
  \label{6.13.2}
  \int\limits_{x\in\mathcal{M}}=L^{\varepsilon}\int_{x}\ ,\quad
  \int_{x}:=\mbox{integral over the torus with $L=1$}\ , 
\end{equation}
such that the overbar in (\ref{6.13.1}) can be thought of as an
averaging procedure, and especially
\begin{equation}
  \label{6.13.3}
  \overline{1}=1\ 
\ .
\end{equation}
Thanks to our regularization prescription the integral of (\ref{6.12})
over internal points can be replaced by $L^{ND}$ (for the
integration measure) times 
\begin{eqnarray}
  \label{6.13}
  \overline{\mbox{Tr} \,\mathfrak{M}}&=&\frac{2N^{2}}{1+N}\ \overline{\C(x_{N{+}1}-x_{i})}\nonumber\\
  &&-\left(\frac{2N(N-1)}{1+N}\right)\left(\overline{\C(x_{N{+}1}-x_{i})}-\frac{1}{2}\
  \overline{\C(x_{i}-x_{j})}\right)\nonumber\\
  &=&\frac{2N}{1+N}\
\overline{\C(x_{N{+}1}-x_{i})}+\frac{N(N-1)}{1+N}\
\overline{\C(x_{i}-x_{j})}\ .
\end{eqnarray}
Due to the internal symmetry of the closed manifolds which we consider
the expression above can be further simplified, since
\begin{equation}
  \label{6.13.1.b}
  \overline{\C(x_{N{+}1}-x_{i})}=\overline{\C(x_{i}-x_{j})}\equiv \overline{\C(x)}\ .
\end{equation}
Introducing a diagrammatic notation
\begin{equation}
  \label{6.14}
  \raisebox{-4.5mm}{\includegraphics[scale=0.09]{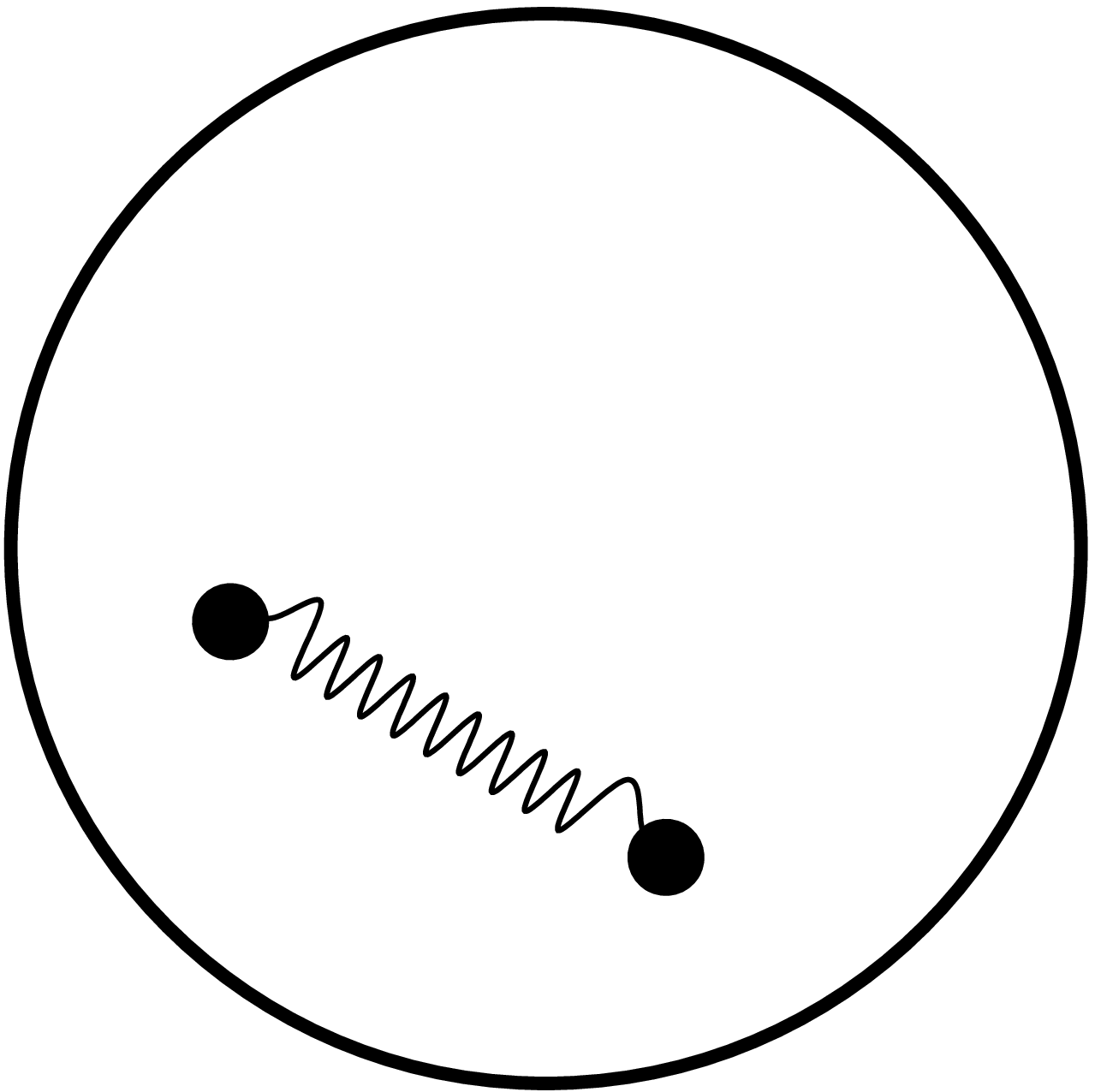}} :=
\overline{\C(x)}\ , 
\end{equation}
the $N$-loop integral reads up to first order in $2{-}D$
\begin{equation}\label{6.16}
  \int (\det D)^{-d/2} =\mu^{-N \varepsilon}
    \left(\frac{1+N}{2^{N}}\right)^{-d/2}\!\!\!\!\!\!\times
\Bigg[1-\frac{d}{2}(2{-}D)\left(N\
      \raisebox{-4.5mm}{\includegraphics[scale=0.09]{./eps/Cijn.eps}}\right)+ O((2{-}D)^{2}) \Bigg]\ .
\end{equation}
For the further analysis we will not only need (\ref{6.13}), but also
the terms appearing to higher order in $2-D$ in (\ref{6.71}). We
derived expressions like (\ref{6.13}) for $\mbox{Tr}^{2}M$ and
$\mbox{Tr}M^{2}$ and all terms up to fourth order in $2-D$ with a
\mathematica-program. It is based on the fact that all terms to appear
in the expansion (\ref{6.71}) are of the form $\mbox{Tr}^{n}M^{m}$ or
products of the latter and therefore can be written as
$\mbox{P}(N)/(N{+}1)^{k}$, where $n,m,k\in \mathbb{N}$ and
$\mbox{P}(N)$ is some polynomial in $N$. It will turn out soon that it
is convenient to expand the polynomial $\mbox{P}(N)$ in terms of the
following base:
\begin{equation}
  \label{6.17}
  \lbrace{1,N,N(N{-}1),N(N{-}1)(N{-}2),\dots,\prod_{j=0}^{k}(N{-}j),\dots\rbrace}\ .
\end{equation}
We obtain:
\begin{eqnarray}
  \label{6.17.1}
\overline{\mbox{Tr}\,\mathfrak{M}^{2}}&=&\frac{-2N(N{-}1)-N(N{-}1)
(N{-}2)}{1{+}N}\   \overline{\C(x)}^{2}\nonumber \\
&& +\frac{2N+3N(N{-}1)+N(N{-}1)(N{-}2)}{1{+}N}\ \overline{\C^{2}(x)}\ ,
\end{eqnarray}
and
\begin{eqnarray}
  \label{6.17.2}
  \overline{\mbox{Tr}^{2} \,\mathfrak{M}}=\frac{4N(N{-}1)+N(N{-}1)(N{-}2)}{1{+}N}\
  \overline{\C(x)}^{2}+\frac{2N}{1{+}N}\ \overline{\C^{2}(x)}\ .
\end{eqnarray}
Diagrammatically, the averages can be rewritten as
\begin{equation}
  \label{6.17.3}
  \raisebox{-4.5mm}{\includegraphics[scale=0.09]{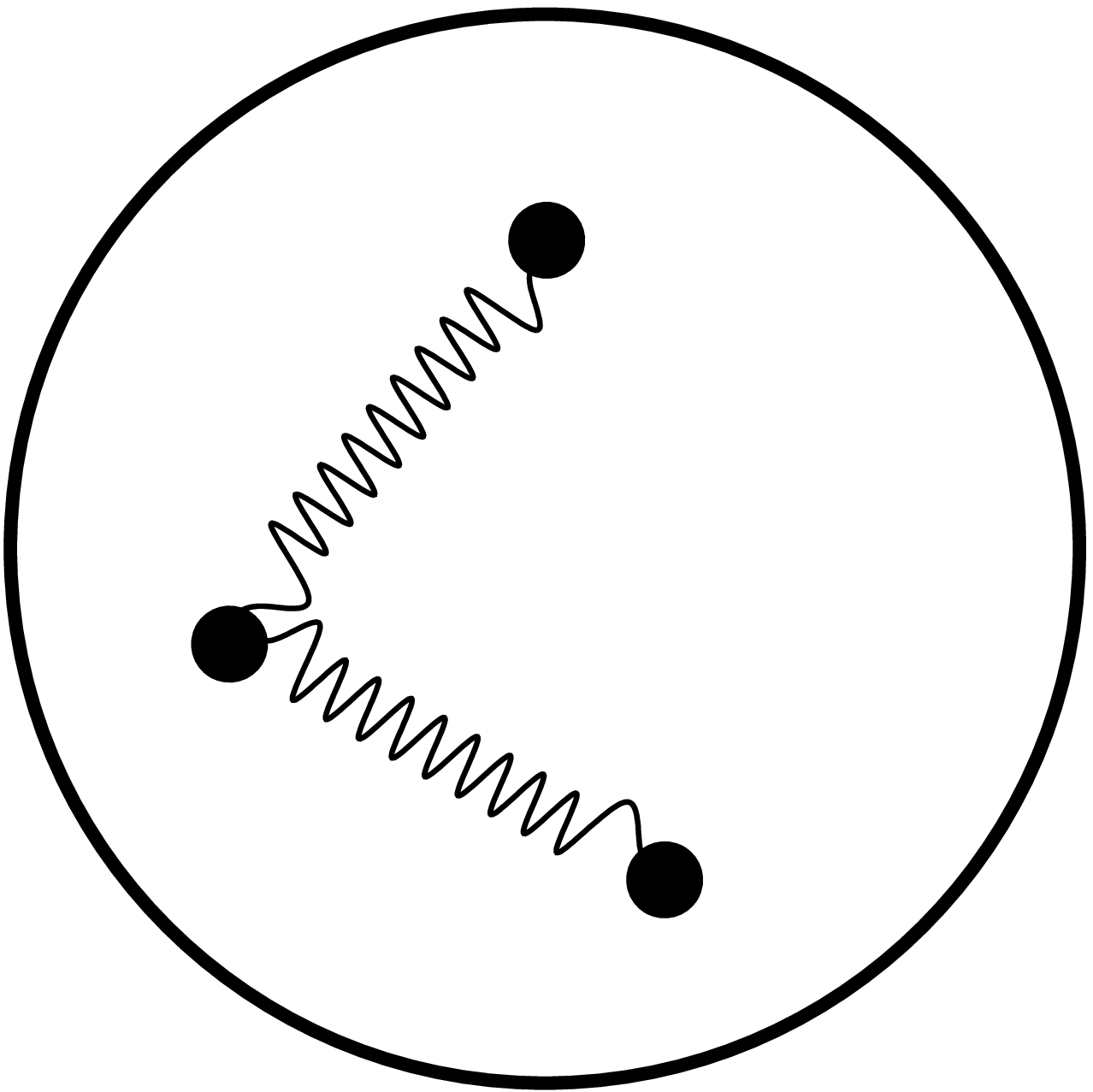}}
  :=\overline{\C(x)}^{2}\ ,
\end{equation}
and
\begin{equation}
  \label{6.17.4}
  \raisebox{-4.5mm}{\includegraphics[scale=0.09]{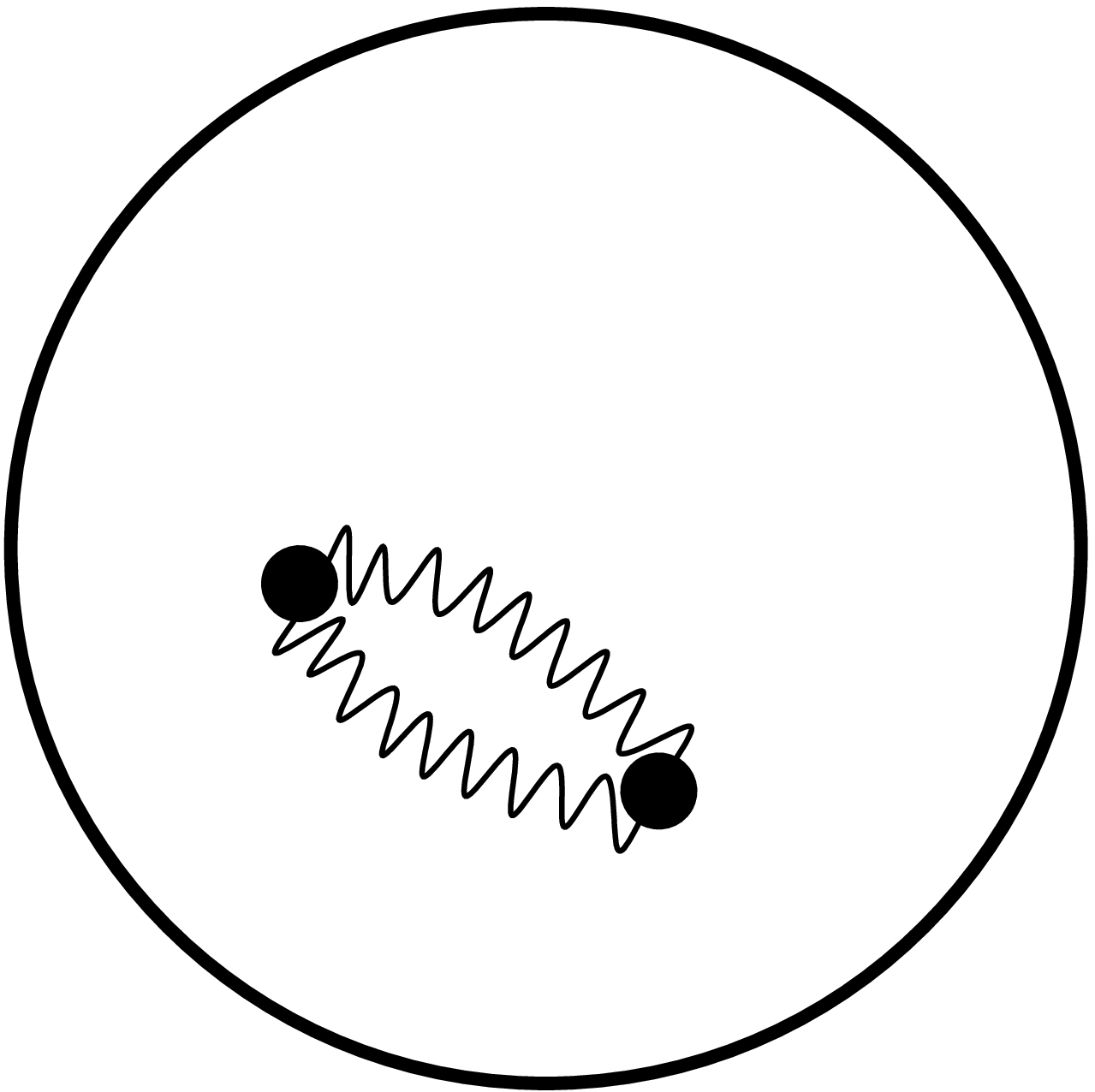}} :=
  \overline{\C^{2}(x)}\ .
\end{equation}
Like in the case of the first order diagram (\ref{6.17.1}) and
(\ref{6.17.2}) are highly simplified as compared to an open manifold,
see our treatment in \cite{PinnowWiese2001}.

Let us shortly discuss the reason for (\ref{6.17}): Inserting
(\ref{6.17.1}) and (\ref{6.17.2}) into the perturbation series and
summing all loop orders, the following series types will appear:
\begin{equation}
  \label{new.0}
  z\sum_{N=0}^{\infty}\frac{\prod_{i=0}^{k{-}1}(N{-}i)(-z)^{N}}{N!(N{+}1)^{d/2{+}j{+}1}}=(-1)^{k}z^{k}\sum_{N=0}^{\infty}\frac{(-z)^{N}}{N!(N{+}k)^{d/2{+}j{+}1}}\equiv
  (-1)^{k}f_{k}^{d{+}2(j{+}1)}(z)\ .
\end{equation}
We may therefore identify the resummed series with a function that we
know already fairly well, in particular we know its strong coupling
behavior. It is furthermore convenient to reduce all functions
$f_{k>1}^{d{+}2(j{+}1)}(z)$ to sums of functions
$f_{1}^{d{+}2(j{+}1)}(z)$ exploiting the formula (\ref{fkdreduce}).

\subsection{Resummed contributions to the expansion in $2-D$ up to
fourth order}\label{2mDresults} We are now almost in the position to
state all resummed contributions up to fourth order in $2-D$. Let us
first state all necessary diagrams:
\begin{eqnarray}
  \label{D1}
  \raisebox{-4.5mm}{\includegraphics[scale=0.09]{./eps/Cijn.eps}} \ =\ 
\overline{\C(x_{i}{-}x_{j})}\ ,
\end{eqnarray}
which contributes to first order in $2-D$. To second order one needs in
addition
\begin{eqnarray}
  \label{D2}
  \raisebox{-4.5mm}{\includegraphics[scale=0.09]{./eps/Cijijn.eps}} \ =\
  \overline{\C^{2}(x_{i}{-}x_{j})}\ .
\end{eqnarray}
To third order diagrams with new topology are
\begin{eqnarray}
  \label{D3}
  \raisebox{-4.5mm}{\includegraphics[scale=0.09]{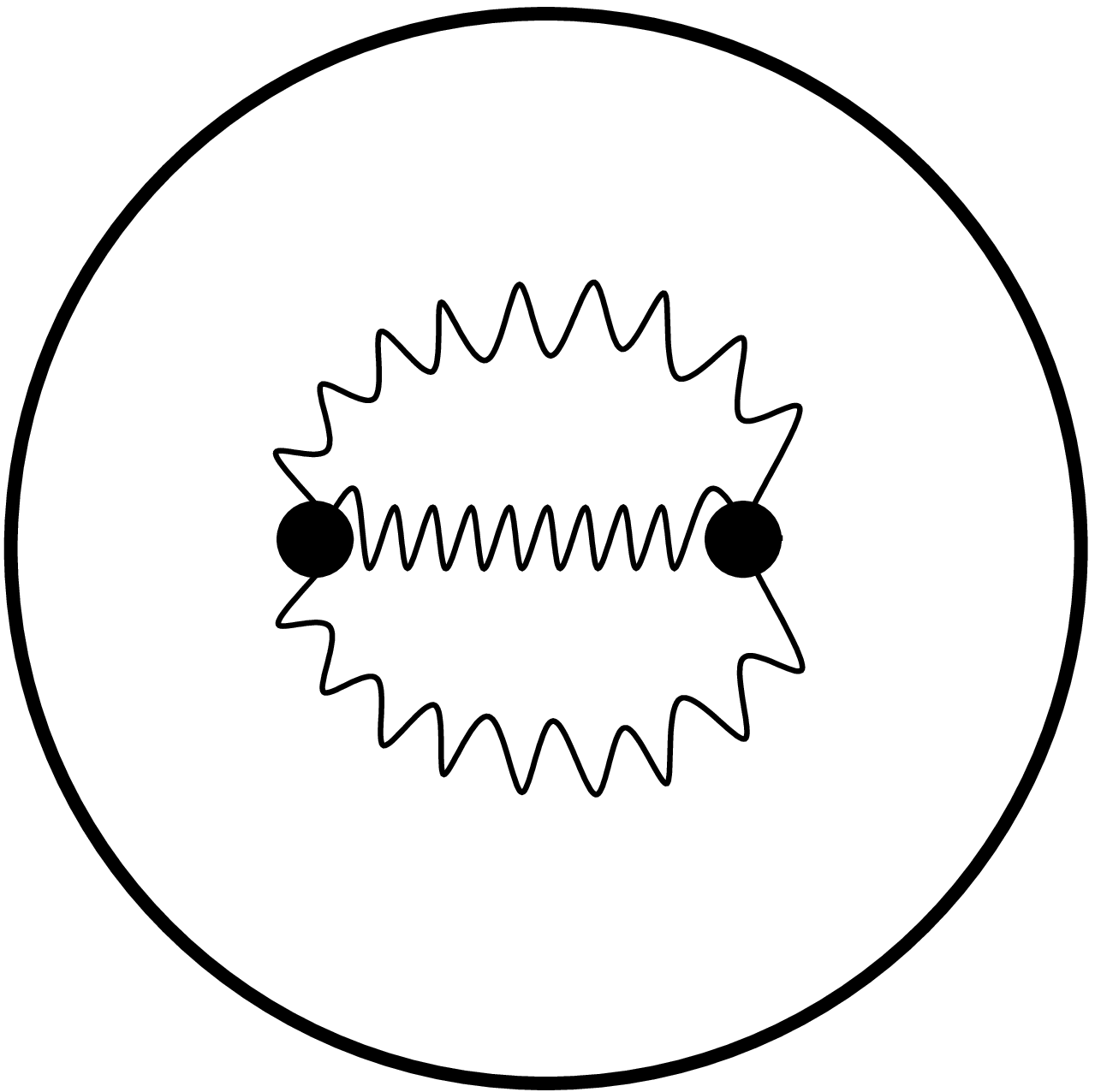}} \ =\
  \overline{\C^{3}(x_{i}{-}x_{j})}\ ,\nn\\
  \raisebox{-4.5mm}{\includegraphics[scale=0.09]{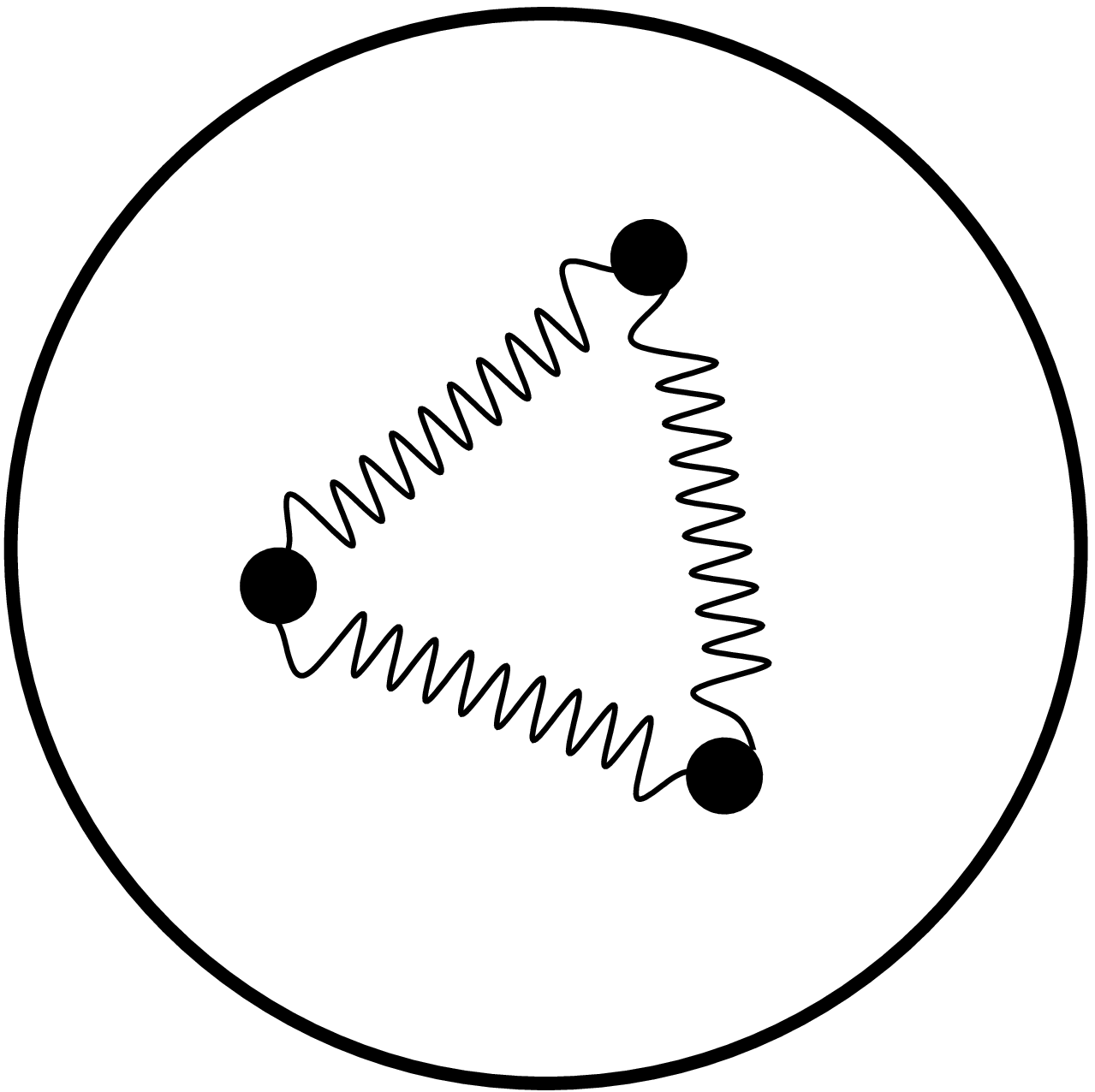}} \ =\
  \overline{\C(x_{i}{-}x_{j})\C(x_{j}{-}x_{k})\C(x_{i}{-}x_{k})}\ .
\end{eqnarray}
Finally, to fourth order arise:
\begin{eqnarray}
  \label{D4}
  \raisebox{-4.5mm}{\includegraphics[scale=0.09]{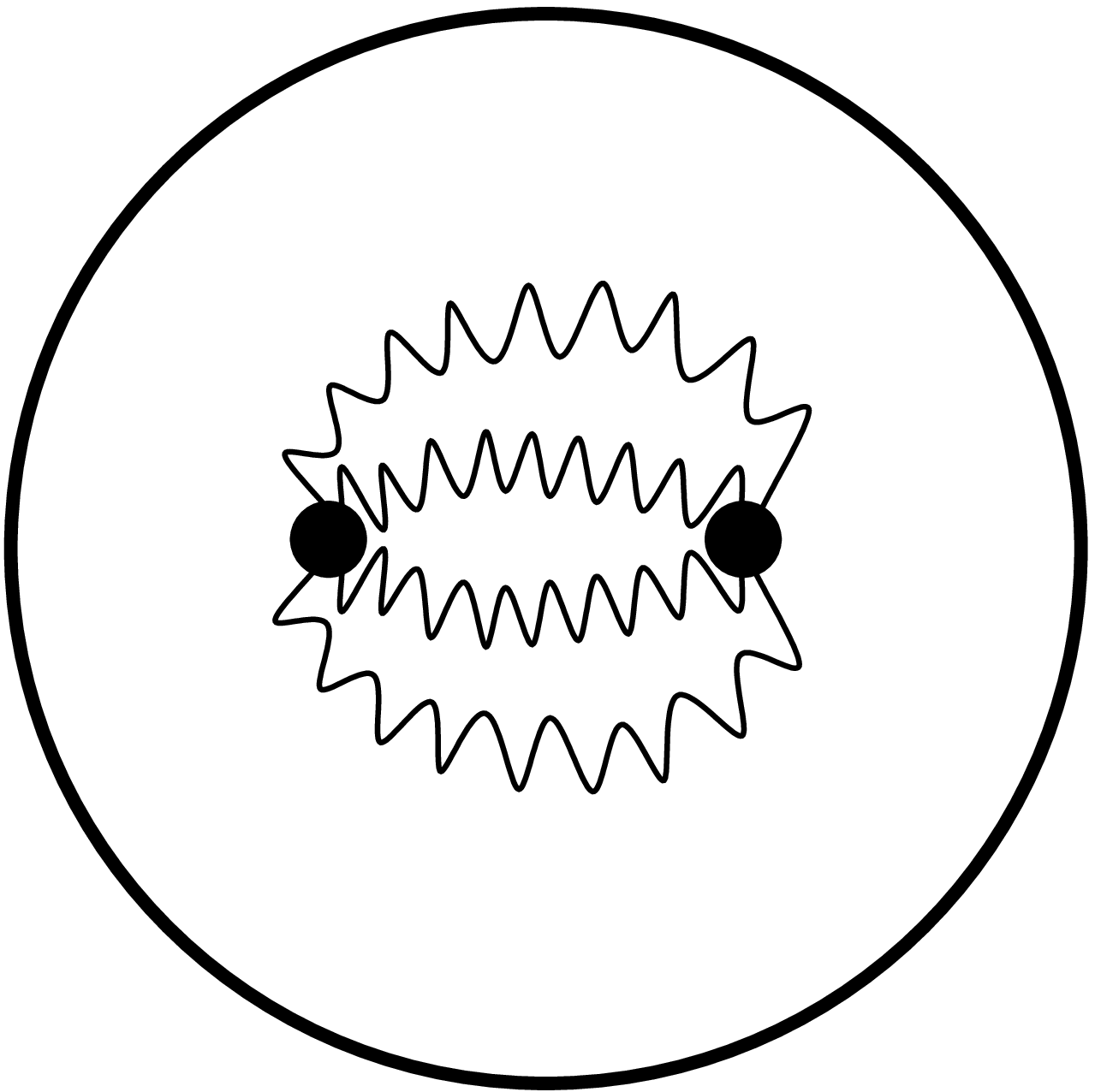}} \ =\
  \overline{\C^{4}(x_{i}{-}x_{j})}\ ,\nn\\
  \raisebox{-4.5mm}{\includegraphics[scale=0.09]{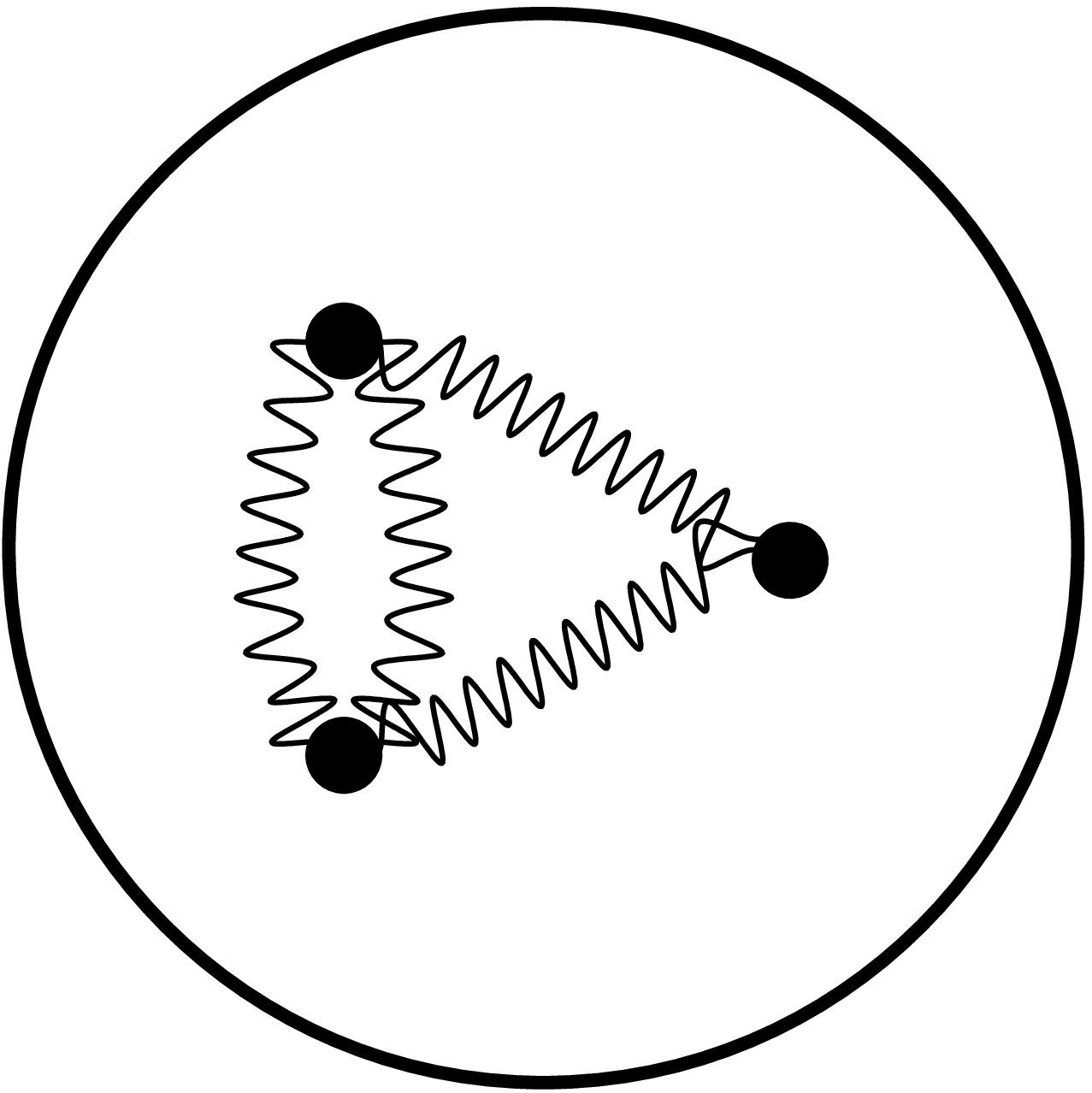}} \ =\
  \overline{\C^{2}(x_{i}{-}x_{j})\C(x_{k}{-}x_{j})\C(x_{k}{-}x_{i})}\ ,\nn\\
  \raisebox{-4.5mm}{\includegraphics[scale=0.09]{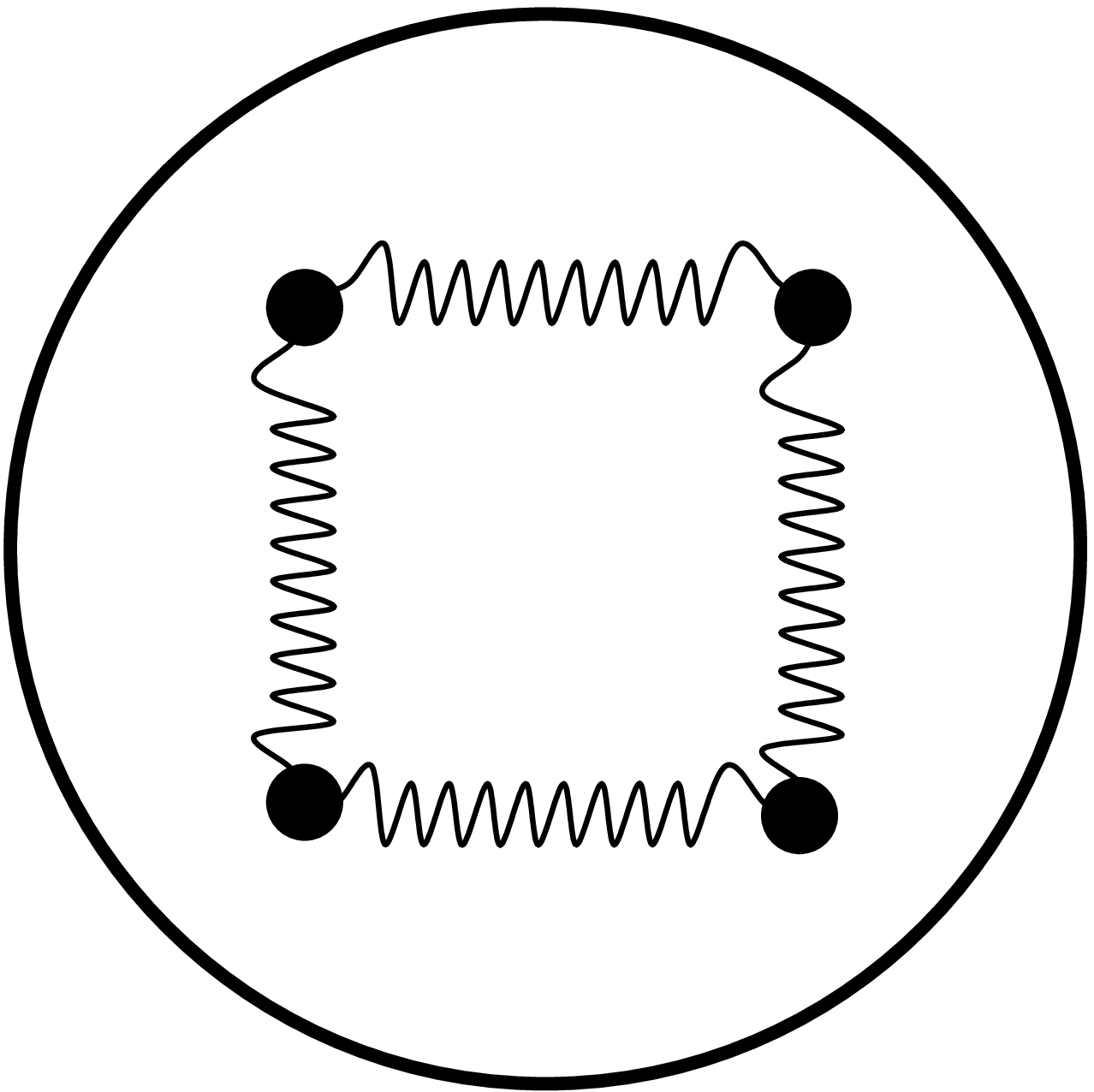}} \ =\
  \overline{\C(x_{i}{-}x_{j})\C(x_{k}{-}x_{j})\C(x_{l}{-}x_{k})\C(x_{i}{-}x_{l})}\ .
\end{eqnarray}
If one calculates diagrams, it will turn out that it is to some extend more
convenient to express the above averages  in terms of averages over a
connected correlation function, which is defined as
\begin{equation}
  \label{conn3}
  \C_{c}(x):=\C(x)-\overline{\C}\ ,
\end{equation}
such that for instance
\begin{equation}
  \label{conn4}
  \overline{\C_{c}^{2}}=\overline{\C^{2}}-\overline{\C}^{2}\ .
\end{equation} 
Furthermore, we will need:
\begin{eqnarray}
  \label{conn5}
\overline{\C_{c}^{3}}=\overline{\C^{3}}-3\overline{\C}\,
\overline{\C^{2}}+2\overline{\C}^{3} \
\end{eqnarray} 
and
\begin{eqnarray}
  \label{conn6}
  \overline{\C_{c}^{\triangle}}&=&\overline{\C_{c}(x_{i}{-}x_{j})\C_{c}(x_{j}{-}x_{k})\C_{c}(x_{k}{-}x_{i})}\nn\\
  &=&\overline{\C(x_{i}{-}x_{j})\C(x_{j}{-}x_{k})\C(x_{k}{-}x_{i})}+3\overline{\C}^{2}\, \overline{\C(x_{i}{-}x_{j})}-3\overline{\C}\, 
\overline{\C(x_{i}{-}x_{j})\C(x_{j}{-}x_{k})}-\overline{\C}^{3}\nn\\
  &=&\overline{\C^{\triangle}}-\overline{\C}^{3}\ ,
\end{eqnarray}
where $x_{i},x_{j},x_{k}$ are distinct points, and the average is over
their positions. In (\ref{conn6}) we exploited the symmetry of the
closed manifold, 
and the definition of $\overline{\C^{\triangle}}$ is
self-evident. Furthermore, we will need to fourth order in $2-D$:
\begin{eqnarray}
  \label{conn7} \Ccbisq=\overline{\C^{4}}+12\overline{\C^{2}}\,
\overline{\C}^{2}-4\overline{\C^{3}}\,
\overline{\C}-3\overline{\C}^{4}\ ,
\end{eqnarray}
\begin{eqnarray}
  \label{conn8}
  \Ccdia&=&\overline{\C_{c}(x_{i}{-}x_{j})\C_{c}(x_{j}{-}x_{k})\C_{c}(x_{k}{-}x_{l})\C_{c}(x_{l}{-}x_{i})}\nn\\
  &=&\overline{\C^{\diamond}}+5\overline{\C}^{4}
\end{eqnarray}
and
\begin{eqnarray}
  \label{conn9}
  \Ccblobt&=&\overline{\C^{2}_{c}(x_{i}{-}x_{j})\C_{c}(x_{i}{-}x_{k})\C_{c}(x_{k}{-}x_{j})}\nn\\
  &=&\overline{\C^{\BT}}-2\overline{\C^{\triangle}}\,
\overline{\C}-\overline{\C^{2}}\, \overline{\C}^{2}+2\overline{\C}^{4}\ .
\end{eqnarray}
Let us now state all terms which appear in the expansion of the
renormalized coupling $g(z)$ up to fourth order in $2-D$ according to
(\ref{6.71}).  We have to calculate at order $N$ of perturbation
theory:
\begin{eqnarray}\label{2mD3}
\overline{\Tr \,\mathfrak{M}} &=& N\, \overline {{\C
}}
\ .
\end{eqnarray}
Inserting this into the perturbation series and summing up the resulting terms
to all orders in $N$ generates the following contributions in the
$2-D$-expansion of the renormalized coupling:
\begin{eqnarray}
  \label{2mD3.01}
  \sum_{N=1}^{\infty}(\det\mathfrak{D}^{(0)})^{-d/2}\frac{\overline{\Tr \,\mathfrak{M}}(-z)^{N{+}1}}{(N{+}1)!}=\overline {{\C}} f_1^{d+2}(z)-\overline {{\C}} f_1^{d}(z)\ ,
\end{eqnarray}
which contributes to first order in $2-D$. 

To second order in $2-D$, we have (\ref{6.17.1}) providing
\begin{eqnarray}
  \label{2mD3.1}
  \sum_{N=1}^{\infty}(\det\mathfrak{D}^{(0)})^{-d/2}\frac{\overline{\Tr \,\mathfrak{M}^{2}}(-z)^{N{+}1}}{(N{+}1)!} &=&2\overline{\C_{c}^2}\
        f_1^{d+4}(z)+(-4\overline{\C_{c}^2}+\overline{\C}^2)f_1^{d+2}(z)\nonumber \\
&&+(-\overline{\C}^{2} +3\overline{\C_{c}^2})\ f_1^d(z)
-\overline{\C_{c}^2}\ f_1^{d-2}(z)\ ,
\end{eqnarray}
and (\ref{6.17.2}) providing
\begin{eqnarray}
  \label{2mD3.2}
  \sum_{N=1}^{\infty}(\det\mathfrak{D}^{(0)})^{-d/2}\frac{\overline{\Tr^{2}\,\mathfrak{M}}(-z)^{N{+}1}}{(N{+}1)!} &=&2\overline{\C_{c}^2}\ f_1^{d+4}(z)-(2\overline{\C_{c}^2}+\overline{\C}^2)f_1^{d+2}(z)\nonumber\\
        &&+2\overline{\C}^2\ f_1^d(z)-\overline{\C}^2\ f_1^{d-2}(z)\
\ .
\end{eqnarray}
Let us now state the terms at third order in $2-D$, which we derived
with the help of a \mathematica- program ($N$ is the loop order.):
\begin{eqnarray}
  \label{2mD8.1}
  \sum_{N=1}^{\infty}(\det\mathfrak{D}^{(0)})^{-d/2}\frac{\overline{\Tr
  \,\mathfrak{M}^{3}}(-z)^{N{+}1}}{(N{+}1)!}
  &=&4(\overline{\C_{c}^{3}}-4\overline{\C_{c}^{\triangle}})f_{1}^{d+4}(z)\nn\\
  &&+(-10\overline{\C_{c}^{3}}+36\overline{\C_{c}^{\triangle}}+6\overline{\C}\,\overline{\C_{c}^{2}})f_{1}^{d+2}(z)\nn\\
  &&+(9\overline{\C_{c}^{3}}-32\overline{\C_{c}^{\triangle}}-12\overline{\C}\,\overline{\C_{c}^{2}}+\overline{\C}^{3})f_{1}^{d}(z)\nn\\
  &&+(3\overline{\C_{c}^{3}}-17\overline{\C_{c}^{\triangle}}-9\overline{\C}\,\overline{\C_{c}^{2}}+\overline{\C}^{3})f_{1}^{d-2}(z)\nn\\
  &&-3(2\overline{\C_{c}^{\triangle}}+\overline{\C}\,\overline{\C_{c}^{2}})f_{1}^{d-4}(z)+\overline{\C_{c}^{\triangle}}\ f_{1}^{d-6}(z)\ ,\qquad 
\end{eqnarray}
\begin{eqnarray}
  \label{2mD8.2}
  \sum_{N=1}^{\infty}(\det\mathfrak{D}^{(0)})^{-d/2}\frac{\overline{\Tr \,\mathfrak{M}\ \Tr
  \,\mathfrak{M}^{2}}(-z)^{N{+}1}}{(N{+}1)!}   
  &=&4(\overline{\C_{c}^{3}}-4\overline{\C_{c}^{\triangle}})f_{1}^{d+6}(z)\nn\\
  &&+(-8\overline{\C_{c}^{3}}+32\overline{\C_{c}^{\triangle}}+2\overline{\C}\,\overline{\C_{c}^{2}})f_{1}^{d+4}(z)\nn\\
  &&+(6\overline{\C_{c}^{3}}-20\overline{\C_{c}^{\triangle}}+2\overline{\C}\,\overline{\C_{c}^{2}}-6\overline{\C}^{3})f_{1}^{d+2}(z)\nn\\
  &&+(-2\overline{\C_{c}^{3}}+4\overline{\C_{c}^{\triangle}}-7\overline{\C}\,\overline{\C_{c}^{2}}+2\overline{\C}^{3})f_{1}^{d}(z)\nn\\
  &&-(\overline{\C}^{3}-4\overline{\C}\,\overline{\C_{c}^{2}})f_{1}^{d-2}(z)-\overline{\C}\,\overline{\C_{c}^{2}}\ f_{1}^{d-4}(z)\ ,\qquad 
\end{eqnarray}
\begin{eqnarray}
  \label{2mD8.3}
  \sum_{N=1}^{\infty}(\det\mathfrak{D}^{(0)})^{-d/2}\frac{\overline{\Tr^3
  \,\mathfrak{M}}(-z)^{N{+}1}}{(N{+}1)!}
  &=&4(\overline{\C_{c}^{3}}-4\overline{\C_{c}^{\triangle}})f_{1}^{d+6}(z)\nn\\
  &&+(-4\overline{\C_{c}^{3}}+24\overline{\C_{c}^{\triangle}}-6\overline{\C}\,\overline{\C_{c}^{2}})f_{1}^{d+4}(z)\nn\\
  &&+(-8\overline{\C_{c}^{\triangle}}+12\overline{\C}\,\overline{\C_{c}^{2}}+\overline{\C}^{3})f_{1}^{d+2}(z)\nn\\
  &&-3(\overline{\C}^{3}+2\overline{\C}\,\overline{\C_{c}^{2}})f_{1}^{d}(z)\nn\\
  &&+3\overline{\C}^{3}f_{1}^{d-2}(z)-\overline{\C}^{3}f_{1}^{d-4}(z)\ .
\end{eqnarray}
To fourth order in $2-D$ we obtain:
\begin{eqnarray}
  \label{2mD13.1}
  &&\hspace{-0.65cm}\sum_{N=1}^{\infty}(\det\mathfrak{D}^{(0)})^{-d/2}\frac{\overline{\Tr^4 \,\mathfrak{M}}(-z)^{N{+}1}}{(N{+}1)!}\nn\\
  &=&8(222\overline{\C}^{4}+6\Csq\Ccsq+3\Ccsq^{2}-\Ccbisq+24\Ccblobt-36\Ccdia)f_1^{d+8}(z)\nonumber\\
  &&+4(804\overline{\C}^4+12\Csq\Ccsq+3\Ccsq^{2}-2\Ccbisq+72\Ccblobt-132\Ccdia-4\ovC\,\Cccub+16\ovC\,\Cctri)f_1^{d+6}(z)\nonumber\\
  &&-4(432\ovC^{4}-3\ovC^{2}\Ccsq-6\Ccsq^{2}+24\Ccblobt-72\Ccdia-8\ovC\,\Cccub+40\ovC\,\Cctri)f_1^{d+4}(z)\nonumber\\
  &&+(287\ovC^{4}-36\ovC^{2}\Ccsq-12\Ccsq^{2}-48\Ccdia-16\ovC\,\Cccub+128\ovC\,\Cctri)f_1^{d+2}(z)\nn\\
  &&+(4\ovC^{4}+36\ovC^{2}\Ccsq-32\ovC\,\Cctri)f_1^{d}(z)\nn\\
  &&+(-6\ovC^{4}-2\ovC^{2}\Ccsq)f_1^{d-2}(z)+4\ovC^{4}f_1^{d-4}(z)-\ovC^{4}f_1^{d-6}(z)
\end{eqnarray}
\begin{eqnarray}
  \label{2mD13.2}
  &&\hspace{-0.65cm}\sum_{N=1}^{\infty}(\det\mathfrak{D}^{(0)})^{-d/2}\frac{\overline{\Tr \,\mathfrak{M}^{2}\
  \Tr^{2}\,\mathfrak{M}}(-z)^{N{+}1}}{(N{+}1)!}\nn\\
  &=&8(222\overline{\C}^{4}+6\Csq\Ccsq+3\Ccsq^{2}-\Ccbisq+24\Ccblobt-36\Ccdia)f_1^{d+8}(z)\nonumber\\
   &&+4(960\ovC^{4}+24\ovC^{2}\Ccsq+\Ccsq^{2}-4\Ccbisq+100\Ccblobt-156\Ccdia)f_1^{d+6}(z)\nonumber\\
  &&-4(714\ovC^{4}+20\ovC^{2}\Ccsq+8\Ccsq^{2}-3\Ccbisq+70\Ccblobt-116\Ccdia-4\ovC\,\Cccub+12\ovC\,\Cctri)f_1^{d+4}(z)\nonumber\\
  &&+(889\ovC^{4}+36\ovC^{2}\Ccsq-2\Ccsq^{2}-28\ovC\,\Cccub-4\Ccbisq+80\Ccblobt-144\Ccdia+88\ovC\,\Cctri)f_1^{d+2}(z)\nn\\
  &&+(-99\ovC^{4}+3\ovC^{2}\Ccsq+8\Ccsq^{2}+16\ovC\,\Cccub-8\Ccblobt+16\Ccdia-48\ovC\,\Cctri)f_1^{d}(z)\nn\\
  &&+(3\ovC^{4}-11\ovC^{2}\Ccsq-2\Ccsq^{2}-4\ovC\,\Cccub+8\ovC\,\Cctri)f_1^{d-2}(z)\nn\\
  &&+(-\ovC^{4}-5\ovC^{2}\Ccsq)f_1^{d-4}(z)-\ovC^{2}\Ccsq f_1^{d-6}(z)
\end{eqnarray}
\begin{eqnarray}
  \label{2mD13.3}
  &&\hspace{-0.65cm}\sum_{N=1}^{\infty}(\det\mathfrak{D}^{(0)})^{-d/2}\frac{\overline{\Tr^2
  \,\mathfrak{M}^{2}}(-z)^{N{+}1}}{(N{+}1)!}\nn\\
  &=&8(222\overline{\C}^{4}+6\Csq\Ccsq+3\Ccsq^{2}-\Ccbisq+24\Ccblobt-36\Ccdia)f_1^{d+8}(z)\nonumber\\
   &&+4(1116\ovC^{4}+36\ovC^{2}\Ccsq+23\Ccsq^{2}-6\Ccbisq+128\Ccblobt-180\Ccdia+4\ovC\,\Cccub-16\ovC\,\Cctri)f_1^{d+6}(z)\nonumber\\
  &&-4(1080\ovC^{4}+47\ovC^{2}\Ccsq+28\Ccsq^{2}-8\Ccbisq+132\Ccblobt-172\Ccdia+8\ovC\,\Cccub-32\ovC\,\Cctri)f_1^{d+4}(z)\nonumber\\
  &&+(2111\ovC^{4}+148\ovC^{2}\Ccsq+44\Ccsq^{2}+24\ovC\,\Cccub-24\Ccbisq+272\Ccblobt-328\Ccdia-80\ovC\,\Cctri)f_1^{d+2}(z)\nn\\
  &&+(-538\ovC^{4}-74\ovC^{2}\Ccsq+10\Ccsq^{2}-8\ovC\,\Cccub+10\Ccbisq-72\Ccblobt+80\Ccdia+8\ovC\,\Cctri)f_1^{d}(z)\nn\\
  &&+(59\ovC^{4}+20\ovC^{2}\Ccsq-15\Ccsq^{2}-2\Ccbisq+8\Ccblobt-8\Ccdia)f_1^{d-2}(z)\nn\\
  &&+(-2\ovC^{2}\Ccsq+6\Ccsq^{2})f_1^{d-4}(z)-\Ccsq^{2}f_1^{d-6}(z)
\end{eqnarray}
\begin{eqnarray} 
  \label{2mD13.4}
  &&\hspace{-0.65cm}\sum_{N=1}^{\infty}(\det\mathfrak{D}^{(0)})^{-d/2}\frac{\overline{\Tr \,\mathfrak{M}\ \Tr
  \,\mathfrak{M}^3}(-z)^{N{+}1}}{(N{+}1)!}\nn\\
  &=&8(222\overline{\C}^{4}+6\Csq\Ccsq+3\Ccsq^{2}-\Ccbisq+24\Ccblobt-36\Ccdia)f_1^{d+8}(z)\nonumber\\
  &&+4(1038\ovC^{4}+30\ovC^{2}\Ccsq+18\Ccsq^{2}-5\Ccbisq+114\Ccblobt-168\Ccdia+2\ovC\,\Cccub-8\ovC\,\Cctri)f_1^{d+6}(z)\nonumber\\
  &&-2(1818\ovC^{4}+54\ovC^{2}\Ccsq+39\Ccsq^{2}+\ovC\,\Cccub-9\Ccbisq+204\Ccblobt-294\Ccdia-22\ovC\,\Cctri)f_1^{d+4}(z)\nonumber\\
  &&+(1583\ovC^{4}+48\ovC^{2}\Ccsq+36\Ccsq^{2}-\ovC\,\Cccub-6\Ccbisq+186\Ccblobt-258\Ccdia+8\ovC\,\Cctri)f_1^{d+2}(z)\nn\\
  &&+(-358\ovC^{4}-21\ovC^{2}\Ccsq-6\Ccsq^{2}+6\ovC\,\Cccub-48\Ccblobt+60\Ccdia-37\ovC\,\Cctri)f_1^{d}(z)\nn\\
  &&+(35\ovC^{4}+12\ovC^{2}\Ccsq-3\ovC\,\Cccub+6\Ccblobt-6\Ccdia+23\ovC\,\Cctri)f_1^{d-2}(z)\nn\\
  &&+(-3\ovC^{2}-7\ovC\,\Cctri)f_1^{d-4}(z)+\ovC\,\Cctri f_1^{d-6}(z)
\end{eqnarray}
\begin{eqnarray}
  \label{2mD13.5}
  &&\hspace{-0.65cm}\sum_{N=1}^{\infty}(\det\mathfrak{D}^{(0)})^{-d/2}\frac{\overline{\Tr \,\mathfrak{M}^4}(-z)^{N{+}1}}{(N{+}1)!}\nn\\
  &=&8(222\overline{\C}^{4}+6\Csq\Ccsq+3\Ccsq^{2}-\Ccbisq+24\Ccblobt-36\Ccdia)f_1^{d+8}(z)\nonumber\\
  &&+4(1116\ovC^{4}+36\ovC^{2}\Ccsq+23\Ccsq^{2}-6\Ccbisq+128\Ccblobt-180\Ccdia+4\ovC\,\Cccub-16\ovC\,\Cctri)f_1^{d+6}(z)\nonumber\\
  &&-4(1110\ovC^{4}+39\ovC^{2}\Ccsq+31\Ccsq^{2}+10\ovC\,\Cccub-7\Ccbisq+136\Ccblobt-178\Ccdia-36\ovC\,\Cctri)f_1^{d+4}(z)\nonumber\\
  &&+(2473\ovC^{4}+72\ovC^{2}\Ccsq+82\Ccsq^{2}+36\ovC\,\Cccub-16\Ccbisq+304\Ccblobt-396\Ccdia-128\ovC\,\Cctri)f_1^{d+2}(z)\nn\\
  &&+(-955\ovC^{4}-12\ovC^{2}\Ccsq-36\Ccsq^{2}-12\ovC\,\Cccub+5\Ccbisq-92\Ccblobt+154\Ccdia+68\ovC\,\Cctri)f_1^{d}(z)\nn\\
  &&+(288\ovC^{4}+12\Ccsq^{2}-\Ccbisq+12\Ccblobt-47\Ccdia-24\ovC\,\Cctri)f_1^{d-2}(z)\nn\\
  &&+(-60\ovC^{4}-2\Ccsq^{2}+10\Ccdia+4\ovC\,\Cctri)f_1^{d-4}(z)+(6\ovC^{4}-\Ccdia)f_1^{d-6}(z)
\ .
\end{eqnarray}

\subsection{Renormalized coupling} Combining (\ref{3.4.8}),
(\ref{6.71}) and the results (\ref{2mD3.01})-(\ref{2mD13.5}) from the
preceeding subsection we may now give the exact renormalized coupling
to fourth order in $2-D$. For the sake of compactness, we introduce a
new notation: Since all series contributions are of the form as stated
in (\ref{new.0}), we introduce vectors $\mathbb{M}$ such that
\begin{eqnarray}
  \label{2mD14.00}
\sum_{N=1}^{\infty}(\det\mathfrak{D}^{(0)})^{-d/2}\frac{\overline{\prod_{i=1}^{l}\left(\Tr
\,\mathfrak{M}^{n_{i}}\right)^{m_{i}}}(-z)^{N}}{(N{+}1)!}&\equiv&\sum_{j=\mbox{\tiny{min}}}^{\mbox{\tiny{max}}}\mathbb{M}_{
\tiny{\left(\begin{array}[]{cccc}
        \!\!m_{1} & \!\!m_{2} & \!\!\cdots & \!\!m_{l}\\
        \!\!n_{1} & \!\!n_{2} & \!\!\cdots & \!\!n_{l}
    \end{array}\!\!\right)}
}f_{1}^{d+2j}(z)\nn\\
  &\equiv&\mathbb{M}^{j}_{
    \tiny{\left(\begin{array}[]{cccc}
        \!\!m_{1} & \!\!m_{2} & \!\!\cdots & \!\!m_{l}\\
        \!\!n_{1} & \!\!n_{2} & \!\!\cdots & \!\!n_{l}
    \end{array}\!\!\right|\left.\begin{array}[]{c}
      \!\! \mbox{max}\\
      \!\! \mbox{min}
    \end{array}\right)}
}f_{1}^{d+2j}(z)\ ,
\end{eqnarray}
where $\mbox{\small{max}}$ and $\mbox{\small{min}}$ are some integers,
and summation over the index $j$ is implicit. Inserting the results
for the resummed series contributions into (\ref{6.71}) we find for
the renormalized coupling to fourth order in $2-D$:
\begin{eqnarray}
  \label{star!!!}
  g(z)&=&f_{1}^{d+2}(z)-(2-D)\ \frac{d}{2}\mathbb{M}^{j}_{
    \tiny{\left(\begin{array}[]{c}   
          1\ \\
          1\ 
    \end{array}\!\!\right|\left.\begin{array}[]{c} 
      \!1\\
      \!0
    \end{array}\right)}
}f_{1}^{d+2j}(z)\nn\\
   &&+(2-D)^{2}\left(\frac{d}{4}\mathbb{M}^{j}_{
    \tiny{\left(\begin{array}[]{c}  
          1\ \\
          2\ 
    \end{array}\!\!\right|\left.\begin{array}[]{c}
      \!\!\!2\\
      \!\!\!-1
    \end{array}\right)}
}f_{1}^{d+2j}(z)+\frac{d^{2}}{8}\mathbb{M}^{j}_{
    \tiny{\left(\begin{array}[]{c}   
          2\ \\
          1\ 
    \end{array}\!\!\right|\left.\begin{array}[]{c}
      \!\!\!2\\
      \!\!\!-1
    \end{array}\right)}
}f_{1}^{d+2j}(z)\right)\nn\\
  &&-(2-D)^{3}\left(\frac{d}{4}\mathbb{M}^{j}_{
    \tiny{\left(\begin{array}[]{c}   
          1\ \\
          3\ 
    \end{array}\!\!\right|\left.\begin{array}[]{c}
      \!\!\!2\\
      \!\!\!-3
    \end{array}\right)}
}f_{1}^{d+2j}(z)+\frac{d^{2}}{8}\mathbb{M}^{j}_{
    \tiny{\left(\begin{array}[]{cc}   
          1 & 1\ \\
          1 & 2\ 
    \end{array}\!\!\right|\left.\begin{array}[]{c}
      \!\!\!3\\
      \!\!\!-2
    \end{array}\right)}
}f_{1}^{d+2j}(z)+\frac{d^{3}}{48}\mathbb{M}^{j}_{
    \tiny{\left(\begin{array}[]{c}   
          3\ \\
          1\ 
    \end{array}\!\!\right|\left.\begin{array}[]{c}
      \!\!\!3\\
      \!\!\!-2
    \end{array}\right)}
}f_{1}^{d+2j}(z)\right)\nn\\
  &&+(2-D)^{4}\left(\frac{d}{8}\mathbb{M}^{j}_{
    \tiny{\left(\begin{array}[]{c}   
          4\ \\
          1\ 
    \end{array}\!\!\right|\left.\begin{array}[]{c}
      \!\!\!4\\
      \!\!\!-3
    \end{array}\right)}
}f_{1}^{d+2j}(z)+\frac{d^{2}}{8}\left(\frac{1}{4}\mathbb{M}^{j}_{
    \tiny{\left(\begin{array}[]{c}   
          2\ \\
          2\ 
    \end{array}\!\!\right|\left.\begin{array}[]{c}
      \!\!\!4\\
      \!\!\!-3
    \end{array}\right)}
}f_{1}^{d+2j}(z)+\frac{2}{3}\mathbb{M}^{j}_{
    \tiny{\left(\begin{array}[]{cc}   
          1 & 1\ \\
          1 & 3\ 
    \end{array}\!\!\right|\left.\begin{array}[]{c}
      \!\!\!4\\
      \!\!\!-3
    \end{array}\right)}
}f_{1}^{d+2j}(z)\right)\right.\nn\\
  &&\hspace{1.8cm}\left.+\frac{d^{3}}{32}\mathbb{M}^{j}_{
    \tiny{\left(\begin{array}[]{cc}   
          2 & 1\ \\
          1 & 2\ 
    \end{array}\!\!\right|\left.\begin{array}[]{c}
      \!\!\!4\\
      \!\!\!-3
    \end{array}\right)}
}f_{1}^{d+2j}(z)+\frac{d^{4}}{384}\mathbb{M}^{j}_{
    \tiny{\left(\begin{array}[]{c}   
          4\ \\
          1\ 
    \end{array}\!\!\right|\left.\begin{array}[]{c}
      \!\!\!4\\
      \!\!\!-3
    \end{array}\right)}
}f_{1}^{d+2j}(z)\right)\ +\ O(2-D)^{5}
\ .
\end{eqnarray}
The vector entries $\mathbb{M}^{j}$ are to be be taken from subsection \ref{2mDresults}.

It is more convenient to discuss instead of $g (z)$ an integral
transform. 
From the expansion of $f_{k}^{d} (z)$, namely 
\begin{equation}
  \label{2mD13.6}
  f_{1}^{d{+}2j}(z)=\frac{z}{\Gamma(\frac{d}{2})}\int_{0}^{\infty}\mbox{d}r\
  r^{d/2+j-1}\exp[-z\mbox{e}^{-r}-r]\ ,
\end{equation} and the structure of the expansion of $g(z)$ in powers
of $2-D$ and the integral representation of the $f_{1}^{d{+}2j}$ it
follows that the exact renormalized coupling can be written as
\begin{eqnarray}
  \label{2mD13.7}
  g(z)\equiv g(D,z)=z\int_{0}^{\infty}\mbox{d}r\
  \tilde g(r)\exp[-z\mbox{e}^{-r}-r]\ ,
\end{eqnarray}
where $\tilde g(r)$ is of the form
\begin{equation}\label{2mD13.7.1} 
\tilde g(r)=r^{d/2}\bigg[\frac{1}{\Gamma(\frac{d{+}2}{2})}+
(2{-}D)\sum\limits_{n=0}^{\infty}\,\sum
\limits_{j=-n_{\mathrm{max}}}^{n} p_{n_{j}}r^{j}(2{-}D)^{n}\bigg]
\ .
\end{equation}

\subsection{Guessing the exact $\tilde g(r)$}
Let us try to gain more information about the power-law behavior in
(\ref{2mD.0}), that is about the expansion in $2{-}D$ of the
correction-to-scaling exponent $\omega$. Power-law behavior forces the
series (\ref{2mD13.7.1}) to turn into some exponentially decaying
function $\tilde g(r)$ as can be seen from the asymptotic form of $g(z)$:
\begin{equation}\label{exact0}
 g (z)\simeq {\cal A} + {\cal B} z^{-\omega/\varepsilon}= {z}
\int\limits_{0}^{\infty}\mbox{d}r\ \mathrm{e}^{-z\ \rme^{-r}-r}
\left({\mathcal{A}}+\frac{\mathcal{B}\,\rme^{-r \omega /\epsilon }}
{{\Gamma(1{+}\frac{\omega}{\varepsilon})}} \right) +O(\mathrm{e}^{-z})\ .
\end{equation}
In order to check the latter equation note that
\begin{eqnarray}
  \label{2mD14.2}
  f^{2}_{1{+}\frac{\omega}{\varepsilon}}(z)&=&z^{1{+}\frac{\omega}{\varepsilon}}\int_{0}^{\infty}\mbox{d}r\ \exp[-z\
  \rme^{-r}-(1{+}\omega/\varepsilon)r]=\Gamma(1{+}\frac{\omega}{\varepsilon})+O(\mathrm{e}^{-z})\nn\\
  \Rightarrow z^{-\omega/\varepsilon}&=&\frac{z}{\Gamma(1{+}\frac{\omega}{\varepsilon})}\int_{0}^{\infty}\mbox{d}r\ \exp[-z\
  \rme^{-r}-(1{+}\omega/\varepsilon)r]+O(\mathrm{e}^{-z})\nn\\
  &=&\frac{z}{\Gamma(1{+}\frac{\omega}{\varepsilon})}\sum_{n=0}^{\infty}\int_{0}^{\infty}\mbox{d}r\
  \frac{(-\omega/\varepsilon\ r)^{n}}{n!}\exp[-z\ \rme^{-r}-r]+O(\mathrm{e}^{-z})\nn\\
  &=&\frac{1}{\Gamma(1{+}\frac{\omega}{\varepsilon})}\sum_{n=0}^{\infty}\frac{(-\omega/\varepsilon)^{n}}{n!}f_{1}^{2(n{+}1)}(z)+O(\mathrm{e}^{-z})\ ,
\end{eqnarray}
where it is understood that $\omega$ is expanded in powers of $2-D$.\\ 
\\
Let us now test a possible form of the exact $\tilde g(r)$. It should
satisfy the following properties:
\begin{itemize}
\item[(i)] In the limit of $D=2$ the exact result
$r^{d/2}/\Gamma(\frac{d{+}2}{2})$ emerges.
\item[(ii)] For $D<2$ the corresponding $g(z)$ has a finite
fixed-point value together with a strong coupling
expansion. Especially, the ansatz should interpolate to the limit
$D{=}1$, which corresponds to a Gaussian polymer closed to form a
ring. The strong coupling expansion of the renormalized coupling of a
closed chain interacting with a $\delta-$potential is easily obtained
from the factorizability of loop integrals in $D=1$ (see for instance
\cite{PinnowWiese2001}). The result is: 
\begin{equation}
\label{exactD1}
g(z)=\varepsilon\left[1+\sum_{n=1}^{\infty}\left(-\frac{1}{\Gamma(\varepsilon)z
}\right)^n\frac{1}{\Gamma(1{-}n\varepsilon)}\right]\
.  \end{equation}
\item[(iii)] It is consistent with the expansion
(\ref{star!!!}).
\end{itemize}
The (non-unique) ansatz is
\begin{equation}
  \label{exact}
  \tilde g(r)=\mathcal{C}\left(\frac{1-{\mathcal S}(D,r)\
  \mbox{e}^{-\frac{\omega}{\varepsilon}r}}{\omega/\varepsilon}\right)^{d/2}\ ,
\end{equation}
where ${\mathcal S} (D,r)$ is analytic in $D=2$ of the form
\begin{equation}
  \label{exact1}
	{\mathcal S}(D,r)=1+\frac{\omega}{\epsilon}r
	\sum_{n=1}^{\infty}{\mathcal S}_{n}(r)(2{-}D)^{n}\ ,
\end{equation}
and each ${\mathcal S}_{n}(r)$ has a Laurent expansion
\begin{equation}
  \label{exact2}
  {\mathcal S}_{n}(r)=\sum_{j=-n_{\tiny{\mbox{min}}}}^{n_{\tiny{\mbox{max}}}}
s_{n,j}\, r^{j}\ .
\end{equation}
Note, that in the limit of $D\to 2$, the expression
 (\ref{exact}) gives $\mathcal{C} r^{d/2}$, while for $D<2$ it yields upon
integration the form (\ref{2mD.0}), ensuring both properties (i) and
(ii). Let us finally check consistency with the expansion (\ref{star!!!}) up to the second order in $2{-}D$: Inserting
\begin{equation}
  \label{exact3}
  \omega/\varepsilon=\omega_{2}(2{-}D)^{2}{+}O(2{-}D)^{3}
\end{equation}
 (the linear term in $(2{-}D)$ has to vanish\footnote{This is due to
the fact that the order $(2-D)$ term in $g (z)$ scales identically in
$z$ as the leading term. Only the order $ (2-D)^{2}$ diverges more
strongly.}) into the ansatz (\ref{exact}) and expanding to second
order in $2{-}D$ provides
\begin{equation}\label{exact4}
 \tilde g(r)=\mathcal{C}\
r^{d/2}\left[1{-}\frac{d}{2}\left({\mathcal S}_{1}(r)(2{-}D)+\left(\frac{\omega_{2}}{2}r-\frac{d-2}{4}{\mathcal
S}_{1} (r)^{2} 
+ {\mathcal S}_{2}(r)\right)(2{-}D)^{2}{+}\cdots\right)\right]\ .
\end{equation}
Explicitly, (\ref{star!!!}) becomes to second order in $2{-}D$
\begin{eqnarray}
  \label{star4}
  g(z)&=&f_{1}^{d+2}(z)-(2-D)\
\frac{d}{2}\left[\overline {{\C}} f_1^{d+2}(z)-\overline {{\C}} f_1^{d}(z)\right]\nn\\ 
  &&+(2-D)^{2}\frac{d}{4}\left[2\overline{\C_{c}^2}\
        f_1^{d+4}(z)+(\overline{\C}^2-4\overline{\C_{c}^2})f_1^{d+2}(z)+(3\overline{\C_{c}^2}-\overline{\C}^{2})\ f_1^d(z)-\overline{\C_{c}^2}\ f_1^{d-2}(z)\right]\nn\\
  &&+(2-D)^{2}\frac{d^{2}}{8}\left[2\overline{\C_{c}^2}\
    f_1^{d+4}(z)-(2\overline{\C_{c}^2}+\overline{\C}^2)f_1^{d+2}(z)+2\overline{\C}^2\ f_1^d(z)-\overline{\C}^2\ f_1^{d-2}(z)\right]\nn\\
  &&+\ O(2-D)^{3}
\ .
\end{eqnarray}
From this, the first coefficients of the $( 2{-}D)$-expansion of
$\tilde g(r)$ are obtained. They read
\begin{eqnarray}
    \label{exact5} 
\!\!\!\!\tilde g(r) \!&=&\! \frac{r^{d/2}}{\Gamma (\frac{d{+}2}{2})}
\left\{ 1+ (2{-}D)\frac{d}{2}\ovC\left(1 {-}\frac{d}{2r}\right)\right.\!
- (2{-}D)^{2}\!
\left[\frac{d}{2}\Ccsq\ r{+}\frac{d}{4}\left(\Csq{-}4\Ccsq\right){-}\frac{d^{2}}{8}\left(2\Ccsq{+}\Csq\right)\right.\!\!\!\nn\\
&&\hphantom{\frac{r^{d/2}}{\Gamma (\frac{d}{2}+1)}[}\left.\left.
{+}\!\left(\frac{d^{2}}{8}\left({-}\Csq{+}3\Ccsq\right){+}\frac{d^{3}}{8}\Csq\right)r^{{-}1}{-}\frac{d^{2}}{8}\left(\frac{d}{2}{-}1\right)\left(\Ccsq{+}\frac{d}{2}\Csq\right)r^{{-}2}\right]\right\}\ .\ 
\end{eqnarray}
Comparing (\ref{exact4}) and (\ref{exact5}), one identifies
$\mathcal{C}=1/\Gamma(\frac{d{+}2}{2})$, $ {\mathcal
S}_{1}=-\overline{\mathbb{C}}(1{-}\frac{d}{2}\frac{1}{r})$ and
$\omega_{2}=2\overline{\mathbb{C}_{c}^{2}}$, where
$\mathbb{C}_{c}(x){:=}\mathbb{C}(x)-\overline{\mathbb{C}}$.  Note that
the terms proportional to $\overline{\mathbb{C}}^{2}$ in ${\cal S}_{2}
(r)$ mostly cancel with ${\cal S}_{1} (r)^{2}$, a sign that the ansatz
catches some structure.
\\
The  diagrams  to  be calculated  at  this order  are
$\overline{\C}$ and $\overline{\C_{c}^{2}}$ (see appendix B). On a manifold of toroidal
shape,  which  is  equivalent  to periodic  boundary  conditions,  two
discrete sums have to be evaluated:
\begin{eqnarray}\label{exact6} 
\overline{\C}&=&\frac{S_{D}}{4\pi^{2}}\bigg[ \sum_{k \in\mathbb{Z}^{D}\!\!,k\not= 0}\frac{1}{{\vec
k}^{2}}-\frac{2\pi}{(2-D)}\bigg]=-0.44956 +
0.3583\, (2-D)+  O(2-D)^{2}\qquad \\   
  \label{exact7}
\overline{\C_{c}^{2}}&=&\frac{S_{D}^{2}}{16\pi^{4}}\sum_{k
\in\mathbb{Z}^{D}\!\!,k\not= 0}\frac{1}{{\vec k}^{4}}=0.152661 +
O(2-D)\ .
\end{eqnarray}
With the results given above, this leads to
\begin{equation}\label{finalomega}
\omega = 2 \epsilon \overline{\C_{c}^{2}} (2-D)^{2} + O (2-D)^{3} =
0.305322\, \epsilon\,(2- D)^{2} + O (2-D)^{3}\ ,  
\end{equation}
which can be compared to the exact result for $D=1$ (polymers):
$\omega =\epsilon$.  As a caveat, note that the above scheme is not
unambiguous in the sense that the second order term proportional to
$r$ in (\ref{exact6}) could in principle either be attributed to
$\omega_{2}$ or ${\cal S}_{2}$. However, any ansatz in (\ref{exact})
will provide an $\omega$, whose expansion starts at least
quadratically in $2-D$. Though (\ref{exact}) is the best ansatz that
could yet be found ensuring properties (i)-(iii), the precise form of
constraints on the scaling function $\mathcal{S}$ remains to be
discussed in order to settle this question.
\\


\section{Conclusion}\label{conclusion} In this work we refined the
analysis of a $D$-dimensional elastic manifold interacting by some
$\delta $-potential with a fixed point in embedding space. Starting
from the perturbation expansion of the effective coupling of the
problem, in a first step, we performed a new calculation using a
modified regularization prescription: Evaluating loop integrals in
fixed space dimension on a manifold of finite size enforced the
introduction of a microscopic cutoff as soon as $D=2$. This way, we
recovered the complete summability of the perturbation theory in this
limit and confirmed the strong coupling behavior as found previously
in an analytic continuation from below $D=2$. In the strong coupling
limit, corresponding to strong repulsion or equivalently to large
membrane sizes, the effective coupling diverges logarithmically as a
function of the bare coupling $z$ yielding a vanishing
correction-to-scaling exponent $\omega$. Analyzing the RG
$\beta$-function we found that it tends to zero at infinite bare
coupling $z$ as $0\leq d<2$. The renormalization group flow then tends
to a fixed point, and the theory becomes scale invariant in this
limit. Due to the logarithmic divergence of the effective coupling,
however, the corresponding zero of the $\beta$-function in terms of
the latter is, too, shifted to infinity. This is a quite remarkable
result showing that the scaling behavior of the system is accessible
only to an all order treatment and deviates qualitatively from any
finite loop expansion, be it within a minimal subtraction scheme or
at finite $\varepsilon$. Especially, the logarithmic growth of the
effective coupling signals the limiting behavior of a scale-invariant
theory.
\\
The result in $D=2$ is completely independent of the regularization
procedure. This does no longer hold true beyond the leading order,
which should be accessible to an expansion in $2-D$. We constructed
its first order in a specific regularization scheme in
\cite{PinnowWiese2001}. While this reproduces qualitatively correctly
the known result in $D=1$, it suffers from a renormalization scheme,
which neglects the boundaries of a finite manifold. We used a hard
cutoff in position space, while working with the infinite $D$-space
correlator. It seems that only in an $\varepsilon$-expansion this
procedure is systematic. \\
Now, in a second step of we have overcome this problem by constructing
the $2-D$-expansion on a manifold of toroidal shape of finite size,
thus imposing periodic boundary conditions on the field. There is no
further infrared cutoff necessary. We have carried out the expansion
of the renormalized coupling up to fourth order in $2-D$, revealing
the general structure of the expansion. It is important to point out
that in considering $g$ as a function of the bare coupling, the limits
$D\to 2$ and strong coupling ($z\to \infty$) can not be
interchanged. While $g$ tends to infinity as $z$ does in $D=2$, we
expect finiteness of this limit as soon as $2-D>0$ and the existence
of a strong coupling expansion as found for polymers ($D=1$). We were
able to guess an exact $g(D,z)$ as a function of $z$ and the internal
dimension $D$, which satisfies these properties and which can be
reconciled with the available expansion in $2-D$ by an appropriate
matching of its free parameters. Though it turned out that due to an
ambiguity in the matching of parameters the precise power-law behavior
of the effective coupling below $D=2$ can not yet be isolated, we
found that for closed manifolds the expansion of $\omega$ in powers of $2-D$ starts at least quadratically as $D<2$.\\
The exponent is closely related to
observables, which are accessible through Monte-Carlo experiments. These are
for instance plaquettes-density functions at the repelling potential on a
membrane avoiding a single point.\\
While results for the pinning problem are interesting on its own, the
main motivation is certainly to obtain a better understanding of
self-avoiding polymerized membranes. Preliminary studies
\cite{PinnowWieseProgress} indicate that this problem can also be
attacked by the methods developed in this work. This would be very
welcome to settle discrepancies between field theoretic results on one
side \cite{WieseDavid1997,DavidWiese1996,DavidWiese1998} and numerical
results (e.g.\ \cite{BowickTravesset2001}) on the other.


\begin{appendix}
\section{The propagator}\label{sec:ker} The regularized difference
correlator is defined as
\begin{equation}
  \label{c.7}
  C_{a}(x)=G_{a}(0)-G_{a}(x)\ ,
\end{equation}
where $G_{a}(x)$ denotes the usual two-point correlator, which is
obtained from\footnote{Strictly speaking, we have to consider the
propagator on the torus, as is done in appendix \ref{app:B}. However
this does not make any difference for the purpose of our argument.}:
\begin{equation}
  \label{c.8}
  G_{a}(x)=\frac{1}{(2\pi)^{D}}\int\mathrm{d}^{D}k\ \frac{\exp[i\vec k\vec
  x-a^{2}k^{2}]}{k^{2}}\ .
\end{equation}
Here, short-wavelength modes are suppressed through a soft cutoff
procedure. Introducing a Schwinger parameterization for the evaluation
of the integral in (\ref{c.8}),
\begin{eqnarray}
  \label{c.9}
  G_{a}(x)&=& \int_{0}^{\infty}\rmd t \int \frac{\rmd^{D}k}{(2\pi)^{D}}
\rme^{- (t +a^{2}) k^{2}} \, \rme^{ikx}\nonumber \\
&=&\frac{1}{(2\sqrt{\pi})^{D}}\int\limits_{0}^{1/a^{2}}\mathrm{d}s\
s^{D/2-1}\mathrm{e}^{-s\frac{x^{2}}{4}}\ ,
\end{eqnarray}
where $s=1/ (t+a^{2})$, we obtain for (\ref{c.7}):
\begin{equation}
  \label{c.10}
  C_{a}(x)=\frac{1}{(2\sqrt{\pi})^{D}}\int\limits_{0}^{1/a^{2}}\mathrm{d}s\
  s^{D/2-1}\left(1-\mathrm{e}^{-s\frac{x^{2}}{4}}\right)\ .
\end{equation}
Further evaluation leads to:

\noindent 
(i) $D=2$:
  \begin{equation}
    \label{c.11}
    C_{a}(x)=\frac{1}{4\pi}\left(\gamma+\Gamma(0,\frac{x^{2}}{4a^{2}})+\ln\frac{x^{2}}{4a^{2}}\right)\overset{x\to \infty}{\longrightarrow}\frac{1}{2\pi}\ln\frac{x}{a}\ .
\end{equation}

\noindent
(ii) $D<2$:
  \begin{eqnarray}
    \label{c.12}
    &&\hspace{-1cm}C_{a}(x)=\frac{|x|^{2{-}D}\Gamma(\frac{D}{2})}{(2{-}D)\ 2\
    \pi^{D/2}}{+}\frac{a^{2{-}D}}{(2{-}D)\ 2^{D{-}1}\pi^{D/2}}\
    \mathrm{e}^{{-}\frac{x^{2}}{4 a^{2}}}{-}\frac{a^{2{-}D}}{(2{-}D)\
    2^{D{-}1}\pi^{D/2}}{-}\frac{|x|^{2{-}D}\Gamma(\frac{D}{2},\frac{x^{2}}{4
    a^{2}})}{(2{-}D)\ 2\ \pi^{D/2}}\nn\\
 &&\overset{x\to
    \infty}{\simeq}\frac{|x|^{2{-}D}}{S_{D}(2{-}D)}{-}\frac{a^{2{-}D}}{(2{-}D)\ 2^{D{-}1}\pi^{D/2}}\ \overset{D\to 2}{\longrightarrow}\ \frac{1}{2\pi}\ln\frac{x}{a}\ .
  \end{eqnarray}
$\Gamma(z,\alpha)$ denotes the incomplete $\Gamma$-function:
\begin{equation}
  \label{c.13}
  \Gamma(z,\alpha)=\int\limits_{\alpha}^{\infty}\mathrm{d}t\
  t^{z-1}\mathrm{e}^{-t}\ .
\end{equation}
Especially:
\begin{equation}
  \label{c.14}
  \lim_{a\to 0}C_{a}(x)=\frac{|x|^{2-D}}{S_{D}(2-D)}
\ ,
\end{equation}
as long as $D<2$.

\section{Calculation of the diagrams in the $(
2-D)$-expansion}\label{app:B} In this section we calculate the
diagrams which appear in the $2-D$ expansion on the torus of size
$L=1$. It turns out that to obtain $\overline{\C}$ and
$\overline{\C_{c}^{2}}$ we need to evaluate two sums over discrete
wave-vectors due to periodic boundary conditions on the torus. Let us
first derive the latter before turning to the explicit
evaluation. Starting from the definition of the difference correlator
$C(x)$,
\begin{equation}
  \label{d.0.1}
  C(x):=G(x)-G(0)\ ,
\end{equation}
where $G(x)$ is the usual two-point correlator, we obtain $C(x)$
through an inverse discrete Fourier-transformation from $G(k)=1/{\vec
k}^{2}$, which reads:
\begin{equation}
  \label{d.0.2}
  C(x)=\sum_{\vec k\not=0}\frac{1}{{\vec k}^{2}}\left(1-\mbox{e}^{i\vec k \vec
  x}\right)\ ,\qquad \vec k ={2\pi}\vec n\ , \ \vec n\in
  \mathbb{Z}\times\mathbb{Z}\backslash\{\vec 0\}\ .
\end{equation}
Performing the averaging procedure
\begin{equation}
  \label{d.0.3}
  \overline{C(x)}=\int_{x}C(x)\ ,
\end{equation}
where $ \int_{x}\mbox{e}^{i\vec k \vec x}=\delta^{D}_{\vec k}$ is to be taken
into account, the calculation of $\overline{C(x)}$ reduces to
\begin{equation}
  \label{d.1}
  \overline{C(x)}=I_{1}:=\sum_{\vec k\not=0}\frac{1}{{\vec k}^{2}}\ ,\qquad
  \vec k={2\pi}\vec n\ ,
\end{equation}
where $\vec k$ is $D$-dimensional, and the indices $n_{i}$ are integer
and running from $-\infty$ to $\infty$, $\vec n=0$ being excluded from
the summation. Of course, in the expansion in powers of $2-D$ we need
an analytic continuation to real values of D. Finally, to obtain
$\overline{\C(x)}$ we have to subtract $\overline{\C^{(0)}(x)}$ from
$\overline{C(x)}$. Due to our normalizations:
\begin{equation}
  \label{d.1.1}
  \overline{\C(x)}=S_{D}\left(\overline{C(x)}-\frac{\overline{\C^{(0)}(x)}}{2\pi(2-D)}\right)\ ,
\end{equation}
where $S_{D}$ denotes the volume of the unit sphere and $\overline{\C^{(0)}(x)}=1$.\\
Turning to $\overline{\C_{c}^{2}(x)}$, we first note that within our
normalizations we have
\begin{equation}
  \label{d.1.2}
  S_{D}^{-2}\,\overline{\C^{2}(x)}=\overline{(C(x)-\C^{(0)}(x)/(2\pi(2{-}D)))^{2}}=\overline{C(x)^{2}}-2\frac{\overline{C(x)}}{2\pi(2{-}D)}+\frac{1}{(2\pi(2{-}D))^{2}}
\end{equation}
and
\begin{eqnarray}
  \label{d.1.3}
  S_{D}^{-2}\,\overline{\C(x)}^{2}=\overline{C(x)}^{2}-2\frac{\overline{C(x)}}{2\pi(2{-}D)}+\frac{1}{(2\pi(2{-}D))^{2}}
\end{eqnarray}
according to (\ref{d.1.1}), such that
\begin{equation}
  \label{d.1.4}
  \overline{\C_{c}^{2}(x)}\equiv
  \overline{\C^{2}(x)}-\overline{\C(x)}^{2}=S_{D}^{2}\left(\overline{C^{2}(x)}-\overline{C(x)}^{2}\right)\ .
\end{equation}
Knowing already the sum to be evaluated to obtain $\overline{C}$, (\ref{d.1}),
what is left is:
\begin{eqnarray}
  \label{d.1.5}
  \overline{C^{2}(x)}&=&\int_{x}C^{2}(x)=\int_{x}\sum_{\vec
  k\not=0}\sum_{\vec p\not=0}\frac{1}{{\vec k}^{2}}\frac{1}{{\vec
p}^{2}} \Big(\mbox{e}^{i\vec k \vec x}-1\Big) 
\Big(\mbox{e}^{i\vec p \vec x}-1\Big)\nn\\
  &=&\sum_{\vec k\not=0}\sum_{\vec p\not=0}\frac{1}{{\vec
k}^{2}}\frac{1}{\vec p^{2}}\left(\delta^{D}_{\vec k{+}\vec
p}-\delta^{D}_{\vec k}-\delta^{D}_{\vec p}+1\right)=\sum_{\vec
k\not=0}\frac{1}{\vec k^{4}}+\left[\sum_{\vec k\not=0}\frac{1}{{\vec
k}^{2}}\right]^{2}\ .
\end{eqnarray}
Therefore,
\begin{equation}
  \label{d.2}
  S_{D}^{-2}\,\overline{\C_{c}^{2}(x)}=I_{2}:=\sum_{\vec k\not=0}\frac{1}{{\vec k}^{4}}\ ,\qquad
  k_{i}= {2\pi}n_{i}\ .
\end{equation}
Let us first calculate $I_{1}$: Introducing a Schwinger
parameterization we have:
\begin{eqnarray}
  \label{d.3}
  I_{1}&=&\frac{1}{(2\pi)^{2}}\sum_{n_{i}=-\infty \atop
  \vec n\not=0}^{\infty}\frac{1}{\vec
  n^{2}}=\frac{1}{(2\pi)^{2}}\sum_{n_{i}=-\infty \atop
  \vec n \not=0}^{\infty}\int\limits_{0}^{\infty}\mbox{d}s\ \mbox{e}^{-s{\vec
  n}^{2}}=\frac{1}{(2\pi)^{2}}\int\limits_{0}^{\infty}\mbox{d}s\ \left[\left(\sum_{n=-\infty}^{\infty}\mbox{e}^{-sn^{2}}\right)^{D}-1\right]\ ,\nonumber\\
\end{eqnarray}
where it is to be noted that the sum in the last line is only
one-dimensional. Furthermore, from now on it is clear, how $I_{1}$ is
analytically continued to real values of $D$.\\
In order to evaluate this sum, we will make use of a Poisson-transformation,
which reads:
\begin{equation}
  \label{d.4}
  \sum_{n=-\infty}^{\infty}\mbox{e}^{-A(n-z/2)^{2}}=\sqrt{\frac{\pi}{A}}\sum_{l=-\infty}^{\infty}\mbox{e}^{-\frac{\pi^{2}l^{2}}{A}+i\pi lz}\ .
\end{equation}
The contribution from $l=0$ is the approximation of the l.h.s.
through a Gaussian integral. Our aim is to calculate the coefficients of the
$2-D$ expansion of $I_{1}$ numerically using some algebraic manipulation
program. Then, the integration interval in (\ref{d.3}) has to be made finite.
This is done as follows: For any $s_{0}>0$ we have
\begin{eqnarray}
  \label{d.5}
  I_{1}&=&\frac{1}{(2\pi)^{2}}\int\limits_{0}^{s_{0}^{-1}}\mbox{d}s\ \left[\left(\sum_{n=-\infty}^{\infty}\mbox{e}^{-sn^{2}}\right)^{D}{-}1\right]+\frac{1}{(2\pi)^{2}}\int\limits_{s_{0}^{-1}}^{\infty}\mbox{d}s\ \left[\left(\sum_{n=-\infty}^{\infty}\mbox{e}^{-sn^{2}}\right)^{D}{-}1\right]\nn\\
  &=&\frac{1}{(2\pi)^{2}}\int\limits_{0}^{s_{0}^{-1}}\mbox{d}s\ \left[\left(\sum_{n=-\infty}^{\infty}\mbox{e}^{-sn^{2}}\right)^{D}{-}1\right]+\frac{1}{(2\pi)^{2}}\int\limits^{s_{0}}_{0}\frac{\mbox{d}s}{s^{2}}\ \left[\left(\sum_{n=-\infty}^{\infty}\mbox{e}^{-n^{2}/s}\right)^{D}{-}1\right]\ .\qquad 
\end{eqnarray}
For any finite $s_{0}>0$, the sum in the r.h.s.\ integral can be
truncated at some finite $n_{\mbox{\tiny{max}}}$ for all $s\in [0,s_{0}]$.
For the first integral (corresponding to small values of $s$) we make use of
the poissonian formula (\ref{d.4}) with $z=0$:
\begin{eqnarray}
  \label{d.6}
  \sum_{n=-\infty}^{\infty}\mbox{e}^{-s n^{2}}=\sqrt{\frac{\pi}{s}}\sum_{l=-\infty}^{\infty}\mathrm{e}^{-\pi^{2}l^{2}/s}\ .
\end{eqnarray}
Inserting this into (\ref{d.5}), the sum in the first integral can be
truncated at some finite $l$ as well, such that one may approximately write:
\begin{eqnarray}
  \label{d.7}
  I_{1}\approx\frac{1}{(2\pi)^{2}}\int\limits_{0}^{s_{0}^{-1}}\mbox{d}s\
  \left[\left(\sqrt{\frac{\pi}{s}}\sum_{l=-l_{\mathrm{\tiny
  max}}}^{l_{\mathrm{\tiny
  max}}}\mathrm{e}^{-\pi^{2}l^{2}/s}\right)^{D}\!\!\!\!{-}1\right]+\frac{1}{(2\pi)^{2}}\int\limits^{s_{0}}_{0}\frac{\mbox{d}s}{s^{2}}\
  \left[\left(\sum_{n=-n_{\mathrm{\tiny max}}}^{n_{\mathrm{\tiny max}}}\mbox{e}^{-n^{2}/s}\right)^{D}\!\!\!\!{-}1\right]\ .\nn\\
\end{eqnarray}
Choosing $s_{0}$ in a way that $l_{\mathrm{\tiny max}}$ can be set equal to
zero the l.h.s.\ integral can be evaluated analytically:
\begin{eqnarray}
  \label{d.8}
  I_{1}\approx\frac{1}{(2\pi)^{2}}\left(\frac{2\pi^{D/2}}{2-D}s_{0}^{D/2-1}-s_{0}^{-1}\right)+\frac{1}{(2\pi)^{2}}\int\limits^{s_{0}}_{0}\frac{\mbox{d}s}{s^{2}}\
  \left[\left(\sum_{n=-n_{\mathrm{\tiny max}}}^{n_{\mathrm{\tiny max}}}\mbox{e}^{-n^{2}/s}\right)^{D}\!\!\!\!{-}1\right]\ .\nn\\
\end{eqnarray}
There is a pole in $2-D$, which can be easily subtracted
expanding the expression in powers of
$2-D$. The pole is
\begin{equation}
  \label{d.9}
  I_{1}=\frac{1}{2\pi(2-D)}+O((2-D)^{0}) .
\ .
\end{equation}
The precision of the machine that we used to evaluate (\ref{d.8}) was
sufficient in a way
that we could select $s_{0}$ from an interval, such that the sum appearing in
the integrand could
be truncated at some finite $n_{\mbox{\tiny max}}$ and the result was independent
from the precise value of $s_{0}$ within the desired order of
accuracy, therefore, justifying the approximation in (\ref{d.7}). Setting for instance $s_{0}=1.9$ and $n_{\mbox{\tiny max}}=20$ we obtain
with {\em Mathematica$^{\bigcirc\hspace{-2.3mm}\mbox{\tiny R}}$\/}\ :
\begin{equation}
  \label{d.10}
  I_{1}=\frac{1}{2\pi(2-D)}-0.715497(1)-0.00457046(1)(2-D)+O((2-D)^{2})\ .
\end{equation}
On the torus we scaled the square root of the volume of the $D$-dimensional
unitsphere into the field. Accordingly, comparing with (\ref{d.1}) and (\ref{d.1.1}) we
then find:
  \begin{equation}
    \label{d.10.1}
    \overline{\C}=-0.44956(1)+0.3583(1)(2-D)+O((2-D)^{2})\ .
  \end{equation}
Let us turn to the evaluation of $I_{2}$ following the same strategy as above.
Again, setting $L=1$ and introducing a Schwinger parameterization leads to:
\begin{eqnarray}
  \label{d.11}
  I_{2}&=&\frac{1}{(2\pi)^{4}}\int\limits_{0}^{\infty}\mbox{d}s\ s\left[\left(\sum_{n=-\infty}^{\infty}\mbox{e}^{-sn^{2}}\right)^{D}\!\!\!\!{-}1\right]\nonumber\\
  &\approx&\frac{1}{(2\pi)^{4}}\int\limits_{0}^{s_{0}^{-1}}\mbox{d}s\ s
  \left[\left(\sqrt{\frac{\pi}{s}}\sum_{l=-l_{\mathrm{\tiny
  max}}}^{l_{\mathrm{\tiny
  max}}}\mathrm{e}^{-\pi^{2}l^{2}/s}\right)^{D}\!\!\!\!{-}1\right]+\frac{1}{(2\pi)^{4}}\int\limits^{s_{0}}_{0}\frac{\mbox{d}s}{s^{3}}\
  \left[\left(\sum_{n=-n_{\mathrm{\tiny max}}}^{n_{\mathrm{\tiny max}}}\mbox{e}^{-n^{2}/s}\right)^{D}\!\!\!\!{-}1\right]\ ,\nn\\
\end{eqnarray}
where we have once again applied the Poisson-transformation (\ref{d.4}) with
$z=0$ on one part of the integration interval and truncated both series at
some finite values $n_{\mbox{\tiny max}}$ and $l_{\mbox{\tiny max}}$.\\
There is no pole in $2-D$. Since $I_{2}$ appears at second order in $2-D$ we
only need its value at $D=2$. $s_{0}$ has to be chosen from an appropriate
interval. Setting $n_{\mbox{\tiny max}}=n_{\mbox{\tiny max}}=10$ and
$s_{0}=1.1$ we obtain with {\em
  Mathematica$^{\bigcirc\hspace{-2.3mm}\mbox{\tiny R}}$\/}\ :
\begin{equation}
  \label{d.14}
  I_{2}=0.00386695(1)+O((2-D))\ ,
\end{equation}
or, due to the rescaling  by $S_{D}^{2}$, 
\begin{equation}
  \label{d.14.1}
  \overline{\C_{c}^{2}}=0.152661(1)+O((2-D))\ .
\end{equation}

\end{appendix}


\section*{Acknowledgments} It is a pleasure to thank R.\ Blossey, F.\
David, H.W.\ Diehl, M.\ Kardar, and L.\ Sch\"afer for useful
discussions. We are grateful to Andreas Ludwig for persisting
questions, and his never tiring efforts to understand the limit of
$D\to 2$.  This work has been supported by the DFG through the Leibniz
program Di 378/2-1, under Heisenberg grant Wi 1932/1-1.

\setcounter{section}{17} 
\section{References}
\def\refname{} \vspace*{-1.2cm}

\begin{thebibliography}{10}

\bibitem{Schaefer}
L.~Sch\"afer,
\newblock {\em Excluded Volume Effects in Polymer Solutions},
\newblock Springer Verlag, Berlin, Heidelberg, 1999.

\bibitem{DesCloizeauxJannink}
J.~des Cloizeaux and G.~Jannink,
\newblock {\em Polymers in Solution, Their Modelling and Structure},
\newblock Clarendon Press, Oxford, 1990.

\bibitem{DeGennes}
P.-G. de~Gennes,
\newblock {\em Scaling concepts in polymer physics},
\newblock Cornell University Press, Ithaca and London, 1979.

\bibitem{Eisenriegler}
E.~Eisenriegler,
\newblock {\em Polymers near surfaces},
\newblock World Scientific, 1993.

\bibitem{Fixman1955}
M.~Fixman,
\newblock {\em Excluded volume in polymer chains},
\newblock J. Chem. Phys. {\bf 23} (1955)   1656--1659.

\bibitem{SchaferWitten1977}
L.~Schafer and T.A. Witten,
\newblock {\em Renormalized field theory of polymer solutions},
\newblock J. Chem. Phys. {\bf 66} (1977)   2121.

\bibitem{DesCloizeaux1981}
J.~des Cloizeaux,
\newblock {\em Polymers in solutions: {Principles} and applications of a direct
  renormalization method},
\newblock J. de Physique {\bf 42} (1981)   635--652.

\bibitem{Edwards1965}
S.F. Edwards,
\newblock {\em The statistical mechanics of polymers with excluded volume},
\newblock Proc. Phys. Soc. {\bf 85} (1965)   613.

\bibitem{DeGennes1972}
P.-G.~De Gennes,
\newblock {\em Exponents for the excluded volume problem as derived by the
  {Wilson} method},
\newblock Phys. Lett. {\bf A 38} (1972)   339--340.

\bibitem{KantorNelson1987a}
Y.~Kantor and D.R. Nelson,
\newblock {\em Crumpling transition in polymerized membranes},
\newblock Phys. Rev. Lett. {\bf 58} (1987)   2774--2777.

\bibitem{KantorNelson1987b}
Y.~Kantor and D.R. Nelson,
\newblock {\em Phase transitions in flexible polymeric surfaces},
\newblock Phys. Rev. {\bf A 36} (1987)   4020--4032.

\bibitem{KantorKardarNelson1986a}
Y.~Kantor, M.~Kardar  and D.R. Nelson,
\newblock {\em Statistical mechanics of tethered surfaces},
\newblock Phys. Rev. Lett. {\bf 57} (1986)   791--795.

\bibitem{KantorKardarNelson1986b}
Y.~Kantor, M.~Kardar  and D.R. Nelson,
\newblock {\em Tethered surfaces: Statics and dynamics},
\newblock Phys. Rev. {\bf A 35} (1987)   3056--3071.

\bibitem{PaczuskiKardarNelson1988}
M.~Paczuski, M.~Kardar  and D.R. Nelson,
\newblock {\em Landau theory of the crumpling transition},
\newblock Phys. Rev. Lett. {\bf 60} (1988)   2638.

\bibitem{PaczuskiKardar1989}
M.~Paczuski and M.~Kardar,
\newblock {\em Renormalization-group analysis of the crumpling transition in
  large $d$},
\newblock Phys. Rev. {\bf A 39} (1989)   6086--6089.

\bibitem{DavidWiese1996}
F.~David and K.J. Wiese,
\newblock {\em Scaling of self-avoiding tethered membranes: 2-loop
  renormalization group results},
\newblock Phys. Rev. Lett. {\bf 76} (1996)   4564.

\bibitem{WieseDavid1997}
K.J. Wiese and F.~David,
\newblock {\em New renormalization group results for scaling of self-avoiding
  tethered membranes},
\newblock Nucl. Phys. {\bf B 487} (1997)   529--632.

\bibitem{WieseHabil}
K.J. Wiese,
\newblock {\em Polymerized membranes, a review}.
\newblock {\em {\em Volume}~19} of {\em Phase Transitions and Critical
  Phenomena}, Acadamic Press, London, 1999.

\bibitem{KardarNelson1987}
M.~Kardar and D.R. Nelson,
\newblock {\em $\varepsilon$ expansions for crumpled manifolds},
\newblock Phys. Rev. Lett. {\bf 58} (1987)   1289 and 2280 E.

\bibitem{AronovitzLubensky1988}
J.A. Aronovitz and T.C. Lubensky,
\newblock {\em Fluctuations of solid membranes},
\newblock Phys. Rev. Lett. {\bf 60} (1988)   2634--2637.

\bibitem{DDG3}
F.~David, B.~Duplantier  and E.~Guitter,
\newblock {\em Renormalization and hyperscaling for self-avoiding manifold
  models},
\newblock Phys. Rev. Lett. {\bf 72} (1994)   311.

\bibitem{DDG4}
F.~David, B.~Duplantier  and E.~Guitter,
\newblock {\em Renormalization theory for the self-avoiding polymerized
  membranes},
\newblock cond-mat\slash {\bf 9702136} (1997).

\bibitem{Hwa1990}
T.~Hwa,
\newblock {\em Generalized $\varepsilon$ expansion for self-avoiding tethered
  manifolds},
\newblock Phys. Rev. {\bf A 41} (1990)   1751--1756.

\bibitem{WieseDavid1995}
K.J. Wiese and F.~David,
\newblock {\em Self-avoiding tethered membranes at the tricritical point},
\newblock Nucl. Phys. {\bf B 450} (1995)   495--557.

\bibitem{ChianelliPrestridgePecoradoPecoradodDeNeufville1979}
R.R. Chianelli, E.B. Prestridge, T.A. Pecorado  and J.P. de~Neufville,
\newblock {\em Molybdenum disulfide in the poorly crystalline ``rag''
  structure},
\newblock Science {\bf 203} (1979)   1105.

\bibitem{HwaKokufutaTanaka1991}
T.~Hwa, E.~Kokufuta  and T.~Tanaka,
\newblock {\em Conformation of graphite oxide membranes in solution},
\newblock Phys. Rev. {\bf A 44} (1991)   2235.

\bibitem{WenGarlandHwaKardarKokufutaLiOrkiszTanaka1992}
X.~Wen, C.W. Garland, T.~Hwa, M.~Kardar, E.~Kokufuta, Y.~Li, M.~Orkisz  and
  T.~Tanaka,
\newblock {\em Crumpled and collapsed conformations in graphite oxide
  membranes},
\newblock Nature {\bf 355} (1992)   426.

\bibitem{SpectorNaranjoChiruvoluZasadzinski1994}
M.S. Spector, E.~Naranjo, S.~Chiruvolu  and J.A. Zasadzinski,
\newblock {\em Conformations of a tethered membrane: Crumpling in graphitic
  oxide?},
\newblock Phys. Rev. Lett. {\bf 73} (1994)   2867--2870.

\bibitem{Baumgaertner1991}
A.~Baumg\"artner,
\newblock {\em Does a polymerized membrane crumple?},
\newblock J. Phys. I France {\bf 1} (1991)   1549--1556.

\bibitem{BaumgaertnerRenz1992}
A.~Baumg\"artner and W.~Renz,
\newblock {\em Crumpled self-avoiding tethered surfaces},
\newblock Europhys. Lett. {\bf 17} (1992)   381--386.

\bibitem{KrollGompper1993}
D.M. Kroll and G.~Gompper,
\newblock {\em Floppy tethered networks},
\newblock J. Phys. I France {\bf 3} (1993)   1131.

\bibitem{BowickTravesset2001}
G.~Thorleifsson M.~Bowick, A.~Cacciuto and A.~Travesset,
\newblock {\em Universality classes of self-avoiding fixed-connectivity
  membranes},
\newblock Eur. Phys. J. {\bf E 5} (2001)   149.

\bibitem{PinnowWiese2001}
H.A. Pinnow and K.J. Wiese,
\newblock {\em Interacting crumpled manifolds},
\newblock J. Phys. A {\bf 35} (2002)   1195--1229.

\bibitem{PinnowWiese2002a}
H.A. Pinnow and K.J. Wiese,
\newblock {\em Interacting crumpled manifolds: Exact results to all orders of
  perturbation theory},
\newblock Europhys. Lett. {\bf 64} (2003)   371--377.

\bibitem{DDG1}
F.~David, B.~Duplantier  and E.~Guitter,
\newblock {\em Renormalization of crumpled manifolds},
\newblock Phys. Rev. Lett. {\bf 70} (1993)   2205.

\bibitem{DDG2}
F.~David, B.~Duplantier  and E.~Guitter,
\newblock {\em Renormalization theory for interacting crumpled manifolds},
\newblock Nucl. Phys. {\bf B 394} (1993)   555--664.

\bibitem{ForgasLipowskyNieuwenhuizenInDombGreen}
G.~Forgas, R.~Lipowsky  and T.M. Nieuwenhuizen,
\newblock {\em The behaviour of interfaces in ordered and disordered systems}.
\newblock {\em {\em Volume}~14} of {\em Phase Transitions and Critical
  Phenomena}, pages 136--376, Academic Press London, 1991.

\bibitem{BrezinHalperinLeibler1983}
E.~Br\'ezin, B.I. Halperin  and S.~Leibler,
\newblock {\em Critical wetting in three dimensions},
\newblock Phys.~Rev. Lett. {\bf 50} (1983)   1387.

\bibitem{Upton1999}
P.J. Upton,
\newblock {\em Exact interface model for wetting in the planar ising model},
\newblock Phys. Rev. E {\bf 60} (1999)   3475--3478.

\bibitem{PinnowWieseProgress}
H.A. Pinnow and K.J. Wiese,
\newblock work in progress.

\bibitem{DavidWiese1998}
F.~David and K.J. Wiese,
\newblock {\em Large orders for self-avoiding membranes},
\newblock Nucl. Phys. {\bf B 535} (1998)   555--595.

\end{thebibliography}

\end{document}